\documentclass[12pt]{iopart}
\usepackage{iopams}  
\usepackage{graphicx}
\usepackage{color}
\usepackage{subfigure}
\usepackage[dvipsnames]{xcolor}

\usepackage{tikz}
\usetikzlibrary{arrows,intersections}
\usepackage{pgfplots}
\usepackage[all]{xy}

\begin{document}

\title{What  do Bloch  Electrons in a Magnetic Field have to do with Apollonian packing of Circles  ?}

\author{Indubala I. Satija}

\address{ Department of Physics, George Mason University, Fairfax, Virginia, USA,\\ 
}
\vspace{10pt}
\ead{
isatija@gmu.edu, 
}
\begin{indented}
\item March 2020
\end{indented}

\begin{abstract}
 Integral Apollonian packing, the packing of circles with integer curvatures, where every circle  is tangent to three other mutually tangent circles, is shown to encode the fractal structure of the energy  spectrum of two-dimensional Bloch electrons in a magnetic field, known as the ``Hofstadter butterfly". In this Apollonian-Butterfly-Connection,  dubbed as $\mathcal{ABC}$,  the integer curvatures of the circles
 contain in a convoluted form,  the topological quantum numbers  of the butterfly graph --  the quanta of the Hall conductivity. Nesting properties of these two fractals are 
  described in terms of the Apollonian group and the conformal  transformations.
  The $\mathcal{ABC}$ unfolds as the conformal maps describing
  butterfly recursions are related to the conformal maps describing nesting of circles in the  Apollonian packing.  Mapping of  butterflies to Apollonian at all scales where Farey tree hierarchy
  plays the central role, reveals how  geometry  and number theory are  intertwined in the quantum mechanics of Bloch electrons in a magnetic field.
\end{abstract}

\section{Introduction}

The ``Hofstadter butterfly" \cite{Hof76,DL} is a quantum fractal representing the energy spectrum of two-dimensional electron gas in a square lattice, subjected to a transverse magnetic field.  It is a physical model for  topological states of matter known as the integer quantum Hall states\cite{Thou}. The ``butterfly graph" is  composed of a nested set of images where each sub-image, although somewhat distorted, is a replica of the original spectrum resembling a butterfly.  Here we show that  the number theoretical aspect of this  fractal is intimately related to an abstract fractal known as  the Apollonian gasket\cite{Apacking,iap}.
The Apollonian gasket describes  the packing of circles  consisting of a nested set of configurations of four mutually tangent  circles.  Every butterfly in the butterfly graph is mapped onto an
integral Apollonian gasket or $\mathcal{IAG}$  - an Apollonian packing  where all  circles have integer curvatures.  This relationship between the butterfly graph and the  $\mathcal{IAG}$  will be referred  as  the {\it Apollonian-Butterfly-Connection} or $\cal{ABC}$. 

\begin{figure}[htbp] 
\includegraphics[width = 1.0\linewidth,height=.8\linewidth]{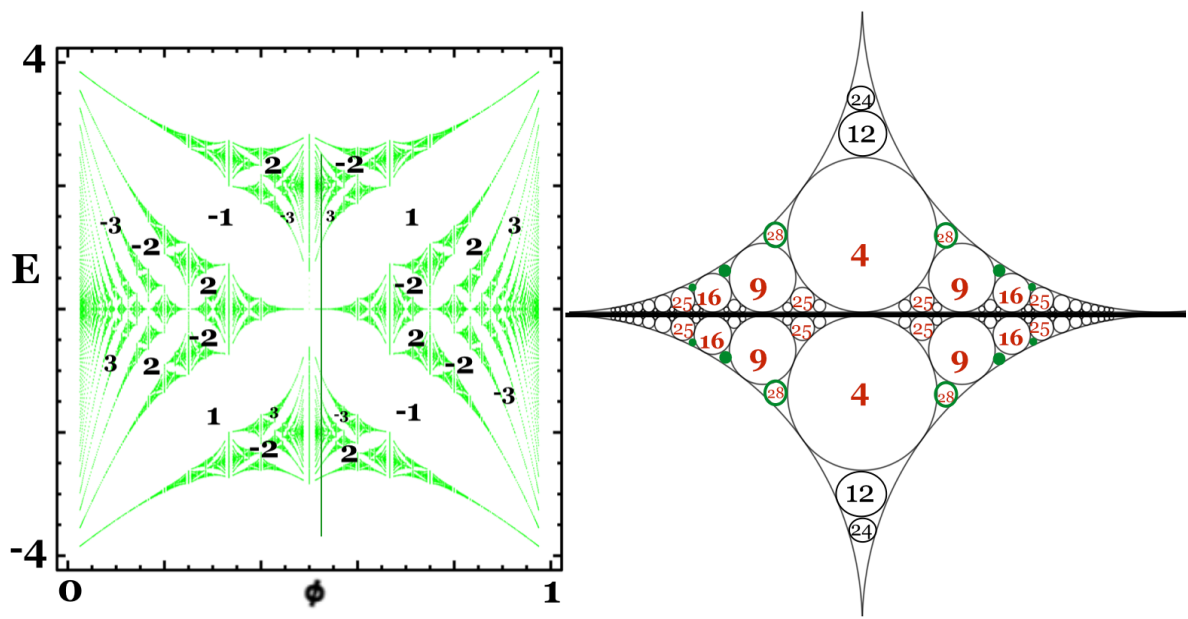}
\leavevmode \caption{ Hofstadter butterfly (left) and $\cal{IAG}$ (right).  The gaps - the forbidden regions of the spectrum are labeled with integers that represent the quantum numbers of Hall conductivity.
In the Apollonian packing, each circle is labeled by its curvature, an integer,  which is the inverse of the radius.}
\label{BP}
\end{figure}

 Fig. (\ref{BP}) shows  the Hofstadter butterfly and the $\cal{IAG}$ -- the two fractals that are labeled by rather  intricate sets of integers.
 In  the butterfly graph,  these integers are topological quantum numbers of Hall conductivity\cite{Thou}.  In $\cal{IAG}$,  the integers  represent  the  integer curvatures of the circles.
 $\mathcal{ABC}$ is a statement about a convoluted relationship between these two sets of integers.
 
Since its discovery, the hierarchical nature of the butterfly spectrum  has been subject of numerous studies\cite{azb64, Wil84,  Wil87, Wil98, Wil00}. 
Recently, it was suggested that the butterfly fractal is related to  $\mathcal{IAG}$\cite{book, EP, Sat18}. This stimulated further investigation of butterfly spectrum\cite{SW}. The highlight of this study was the
 proof of  the ``Farey relation"  -- the heart of $\mathcal{ABC}$, that were known only empirically before. The Farey relation ( Eq. (\ref{FR1}) ) relates the magnetic flux values at the center and the boundaries of the butterfly
 and provides perhaps the first inkling that number theory is intertwined with the quantum mechanics of Bloch electrons in a magnetic field.

 In this paper, the mathematical  framework that relates the butterfly spectrum with the $\mathcal{IAG}$   is put on a solid foundation using the Apollonian group  and the conformal transformation. The self-similar scaling properties of the  fractals are described in terms of  the eigenvalues of the four generators of the Apollonian group and the fixed points of the conformal transformations.
 The transformations describing the nesting structure of the butterfly  graph and  the corresponding conformal transformations describing the recursive patterns of mutually tangent circles  in $\mathcal{IAG}$ are shown to be related. Although of very different origins,  the two sets of transformations results  in the same scaling properties that characterize the self-similar hierarchical  properties of the two fractals and thus  providing a direct link between the butterfly graph and the $\mathcal{IAG}$.

In section $2$, we will begin with a brief review of the  relevant features of the butterfly spectrum.  Section $3$ describes the  Farey relation that relates the left-edge, the right-edge and the center of every sub-butterfly in the graph.  Section $4$ introduces the `` butterfly quadruplets" - the four integers that uniquely specify the number theoretical address of every sub-butterfly in the butterfly graph. Here we also
discuss the division of  the butterfly graph into the ``C-cell" ( central part ) and the "E-cell" ( edge part). Section $5$ reviews the butterfly recursions as described in Ref. (\cite{SW}). In section $6$, 
we revisit the butterfly recursions, expressing them as conformal transformations.

  Section $7$ describes  $\mathcal{IAG}$ and introduces various equivalent forms of  ``Apollonian quadruplets". The Apollonian group and its four generators  that describes the packing of circles as well as the corresponding conformal transformations are described in section $8$ .  Section $9$ illustrates the C-cell and the E-cell division of the Apollonian packing.
  
  In section $10$, we  discuss the $\cal{ABC}$. This section is divided into various sub-sections starting with the so called Ford circles that describe {\it Central} butterflies -- butterflies that exhibit reflection symmetry. This is followed by discussion on conformal images of Ford circles that describe the {\it Edge}-butterflies -- that reside at the  edge of the spectrum.  It also shows the link between conformal transformations for butterfly nesting and the corresponding conformal transformation for nested set of circles in $\mathcal{IAG}$. Section $11$ discusses  the $\mathcal{ABC}$ for the butterflies that share the same flux interval, referred as the {\it butterfly siblings}.
  
In Appendix $1$, we list  identities that relate various integers characterizing the butterflies.  Appendix $2$ gives a brief review of the the  circle inversion that is jewel of the Apollonian packing.  Appendix $3$
 describes the super-Apollonian group, that is essential
 for complete correspondence between the butterfly and the Apollonian.
 In Appendix $4$, we review the tree of the Pythagorean triplets  that encodes the hierarchical aspects of  the C-cell  butterflies as described in our earlier study\cite{Sat18}.

 \section{The Hofstadter Butterfly}
\label{sec: 3}
 
 The model system underlying  the butterfly graph is the  ``butterfly Hamiltonian" that describes a simple model of spineless electrons in a two-dimensional square lattice where electrons can hop only to their neighboring sites.  When subjected to a magnetic field, the Hamiltonian is given by,
 
  \begin{equation*}
 H =  \cos a (k_x -\frac{e}{c}A_x) + \cos a( k_y  - \frac{e}{c} A_y),
 \end{equation*}
 where the magnetic field $\vec{B} = \nabla \times \vec{A}$.  Here $a$ is the lattice constant and $H$ is defined in units of the strength of the nearest-neighbor hopping parameter.
 With the choice of Landau gauge,  $(A_x = 0, A_y = Bx )$, and the wave function $\Psi_{n,m} = e^{i k_y m} \psi_n(k_y)$, the problem effectively  reduces to a one-dimensional system of equations.  Known as  the Harper's equation,\cite{Harper} it is given by,
 
  \begin{equation}
 \psi_{n+1}+\psi_{n-1} + 2 \cos( 2 \pi n \phi + k_y) \psi_n = E \psi_n.
 \end{equation}
 
 The parameter $\phi = \frac{B a^2}{ \hbar/e}$ is the magnetic flux per unit cell of the lattice, measured in the unit of flux quanta $\hbar/e$.  The Hamiltonian can also be written as\cite{Wil87},
 
 \begin{equation}
 H = \cos x + \cos p, \,\,\,\, \ [x, p] = i \phi
 \end{equation}
 
 That is, the butterfly graph lives in a space of energy E and ( effective) Planck's constant $\phi$.  The corresponding continuum Hamiltonian $H = p^2+ x^2$ describes the spectrum given by the
 Landau levels\cite{QEbook}. 
 
 For rational values of the flux  $\phi=p/q$, the allowed energies consist of q bands, with the central pair touching when q is even.  
  The butterfly graph describes all possible integer quantum Hall states  of the non-interacting fermions in two dimension.  Characterizing every gap  of the butterfly graph is  an integer that represents the quantum Hall conductance corresponding to the Fermi level in that gap.
 In $1983$  David Thouless\cite{Thou} along with his collaborators showed that  the Hall conductivity  $C_H$ can be written as,

\begin{equation}
C_H =  [\frac{i}{2 \pi} \sum_{n=1}^{n_f}  \int_{T} \{\partial_{k_x}  \psi_n^*  \partial_{ k_y}  \psi_n- \partial_{k_x}  \psi_n  \partial_{ k_y} \psi_n^*\} \, dk_x \, dk_y] \,\ \frac{e^2}{h} \equiv \sigma \frac{e^2}{h}
\label{BC}
\end{equation}

Here $n_{f}$ represents the number of filled bands as the Fermi energy energy lies in the gaps of the spectrum. The quantity in the square bracket can assume only integral values,  denoted as $\sigma$. Thia is the Chern number - a topological quantum number of Hall conductivity.
The integers $\sigma$ can also be obtained as  solutions of a Diophantine equation\cite{Dana}, namely
\begin{equation}
r = p \sigma + q \tau, \, \, \  \rho = \sigma \phi + \tau,
\label{DE}
\end{equation}

where $r$ labels the $r$th gap of the spectrum and  $\rho=\frac{r}{q}$ represents the electron density. The physical significance of $\tau$ remains unclear.
Every butterfly is characterized by $(\sigma_+, \sigma_-)$, the absolute values of the Hall conductivity characterizing the quantum numbers of the two majors gaps of the butterfly.

In addition to $(\sigma, \tau)$ that labels the gap, we can also define the quantum numbers $(M,N)$ associated with a band at flux $\phi = \frac{p}{q}$, given by the following Diophantine equation,

\begin{equation}
M q + N p = 1.
\label{mn}
\end{equation}

Here $ M= \tau_{\uparrow}-\tau_{\downarrow}$ and $ N = \sigma_{\uparrow}-\sigma_{\downarrow}$ where  $ \sigma_{\uparrow} (\sigma_{\downarrow})$ and
 $ \tau_{\uparrow} (\tau_{\downarrow})$  are the topological quantum numbers of
 the upper (lower)
gaps that sandwich the band at $\phi = \frac{p}{q}$. 

       \begin{figure}[htbp] 
\includegraphics[width = .85 \linewidth,height=1.1 \linewidth]{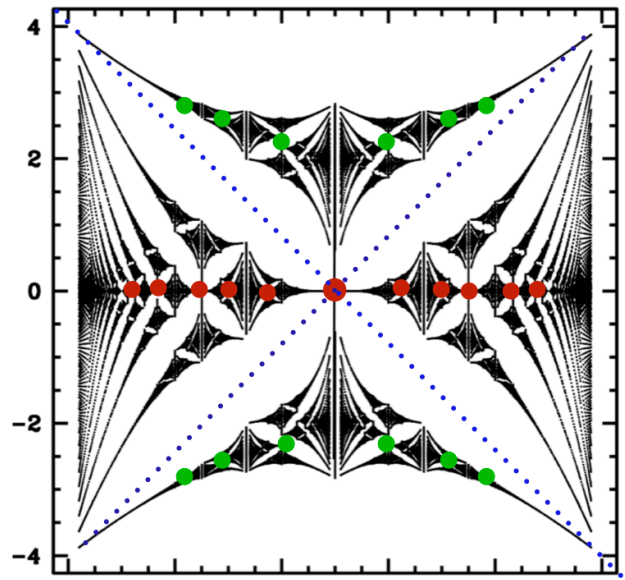}
\leavevmode \caption{Figure illustrates the {\it Central} and the {\it Edge} partitioning  of the butterfly, labeled respectively with the red and the green dots at the centers of the sub-butterflies.  This separation
is due to the two major gaps that are highlighted with dotted lines. }
\label{CEnew}
\end{figure}

\subsection{  C-cell  and E-cell butterflies }

As shown in Fig. (\ref{CEnew}), the major gaps of the butterfly divide the graph into ``Central" and "Edge" parts
 which we will refer as  the C-cell and the  E-cell of the butterfly as shown in Fig. (\ref{CEnew}).  The the E-cells
consist of the upper and the lower parts. Throughout this paper,  the C and the E-cells will  be marked in red  and and  green respectively.  This color coding may be a ``dot" at the  centers of the butterfly, or in some cases the entire butterfly will be color coded

We note that  the C and the E- cells of the main butterfly are ``special" as they have additional symmetries. For example,  every sub-butterfly in the C-cell of the main butterfly has its center at $E=0$ and exhibit mirror symmetry and the entire chain of these butterflies reside symmetrically  ( left and right chains ) about the center of the butterfly graph.  Although  {\it Edge}-butterflies do not exhibit mirror 
symmetries, the upper and the lower E-cells  are mirror image of each other. In our discussion below, the term {\it Central} and {\it Edge} butterflies will be referred to only the C and the E-cell butterflies of the main butterfly.

Below we summarize two important characteristics of the C-cell  and E-cell butterflies which have been proven for the the special case of the {\it Central} and the {\it Edge} hierarchies.
For the general case of C-cell and the E-cell butterflies, these results are empirical, based on observations\cite{Sat18}.\\

(1) The C-cell butterfly hierarchies conserve parity which we define as even (odd) when $q_c$ is even (odd). That is, as we zoom into the equivalent butterflies, $q_c$ retains its even  or odd-ness . We note that butterflies with center at E = 0 are special case of the C-cell butterflies that exhibit even parity and horizontal mirror symmetry at all scales. 
The E-cell butterfly hierarchies do not conserve parity. \\

(2) For C-cell butterfly hierarchies,   $\Delta \sigma \equiv (\sigma_+ - \sigma_-)$ remains constant, equal to $0$ and $1$. In contrast, for  the E-cell butterflies,  $\Delta \sigma$ grows exponentially.
This leads to an asymptotic symmetry in the C cell sub-images,  in sharp contrast to the E cell butterflies that remain asymmetrical.\\

We note that, in addition to the C and  the E-cell butterflies,  there are also  inter-cell butterflies, butterflies whose one wing is the C-cell and another in the E-cell.
However, we will not discuss these images as they are already part of the C and E cell structures.

   \begin{figure}[htbp] 
\includegraphics[width = 0.9\linewidth,height=.7\linewidth]{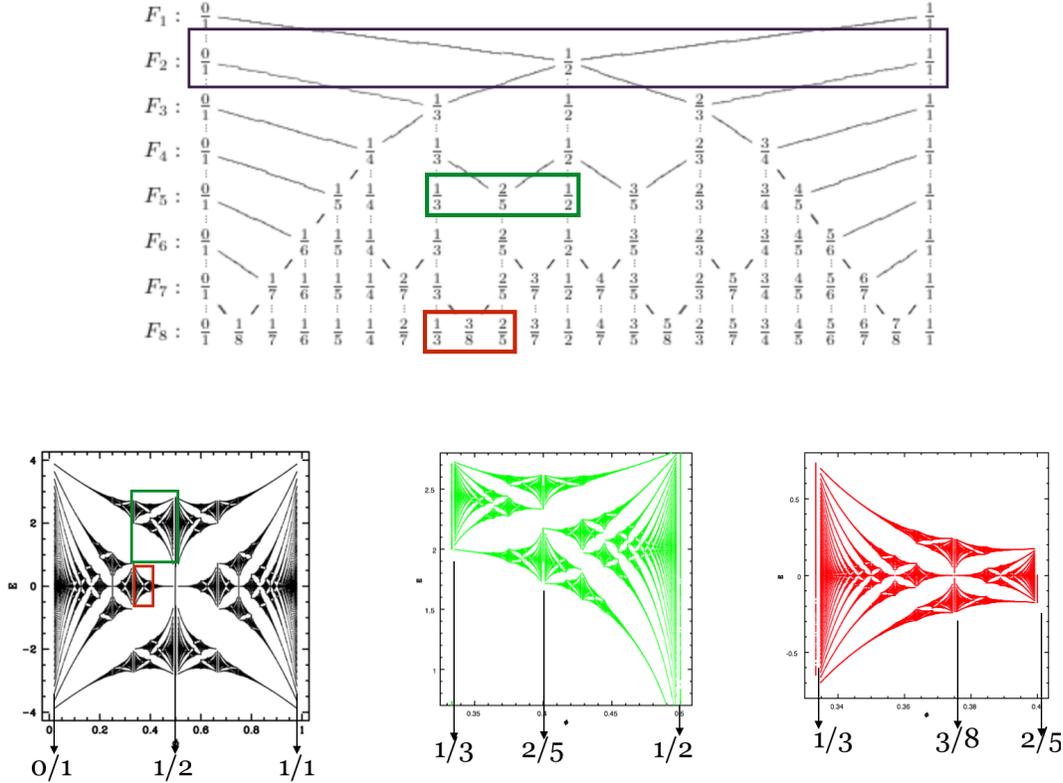}
\leavevmode \caption{ Upper panel  shows the Farey tree: the tree is built up row by row. 
Each successive row of the tree inherits all the Farey fractions from the level above it, and  some new fractions (all of which lie between 0 and 1) are added by combining certain neighbors in the preceding row using an operation called “Farey addition. Some examples of friendly triplets are marked in  rectangular boxes. The bottom part shows
  the corresponding butterflies, all color coded where each Farey triplet forms the flux boundaries and center of a butterfly. }
\label{Farey}
\end{figure}

\section{ Farey Tree and the Butterfly Fractal}

As we stare at the butterfly graph, we see 
various sub-images. These  ``sub-butterflies" that are distorted replicas of the original graph.  Each butterfly with left and right flux boundaries at
$\phi_L=\frac{p_{\rm L}}{q_{\rm L} }$, $\phi_R=\frac{p_{\rm R}}{q_{\rm R}}$ and center at $\phi_c = \frac{p_{\rm c}}{q_{\rm c}}$  has been shown\cite{Wil87,SW} to be  the  renormalization of the original butterfly graph with the corresponding boundaries at $0$ and $1$ and  center at $\frac{1}{2}$. 

Intriguingly, the magnetic flux values at the butterfly boundaries and center are related as,

\begin{equation}
\frac{p_{\rm c}}{q_{\rm c}}=\frac{p_{\rm L}+p_{\rm R}}{q_{\rm L}+q_{\rm R}}
\ .
\label{FR1}
\end{equation}

This relation where 
$\phi_{\rm c}$ is the \lq Farey sum'  of $\phi_{\rm L}$ and $\phi_{\rm R}$, was originally found empirically\cite{book, EP}. It  has been now proven\cite{SW} using the RG equations describing butterfly recursions.
Consequently, the \lq butterfly triplets' $[\frac{p_{\rm L}}{q_{\rm L}}, \frac{p_{\rm c}}{q_{\rm c}}, \frac{p_{\rm R}}{q_{\rm R}}]$  
which are the flux values at the edges and at the 
centers of the sub-images are \lq neighbors'  in the 
Farey tree, known as the \emph{friendly numbers}, satisfying the equations,

\begin{equation}
|q_{\rm L} p_{\rm R} - q_{\rm R} p_{\rm L} | = 1,\,\,\ |q_{\rm L} p_{\rm c} - q_{\rm c} p_{\rm L} | 
= 1,\,\,\ |q_{\rm R} p_{\rm c} - q_{\rm c} p_{\rm R} | = 1
\ .
\label{FR}
\end{equation}

Therefore, the butterfly flux interval $| \Delta \phi |$ - the horizontal size of the butterfly is given by,

\begin{equation}
\Delta \phi = | \frac{p_R}{q_R}-\frac{p_L}{q_L}| = \frac{1}{q_L q_R}
\label{bsize}
\end{equation}

The Farey relation' ( Eq. (\ref{FR1} -\ref{FR}) ) which will emerge as  the key factor underlying the $\mathcal{ABC}$,
 suggests that  number theory  plays an important role in the problem of  Block electrons in a magnetic field. 
Fig. (\ref{Farey}) shows the Farey tree\cite{Hardy} and illustrates the Farey relation (\ref{FR1}) in various sub-butterflies of the butterfly graph.   

    \begin{figure}[htbp] 
\includegraphics[width = .8 \linewidth,height=.65\linewidth]{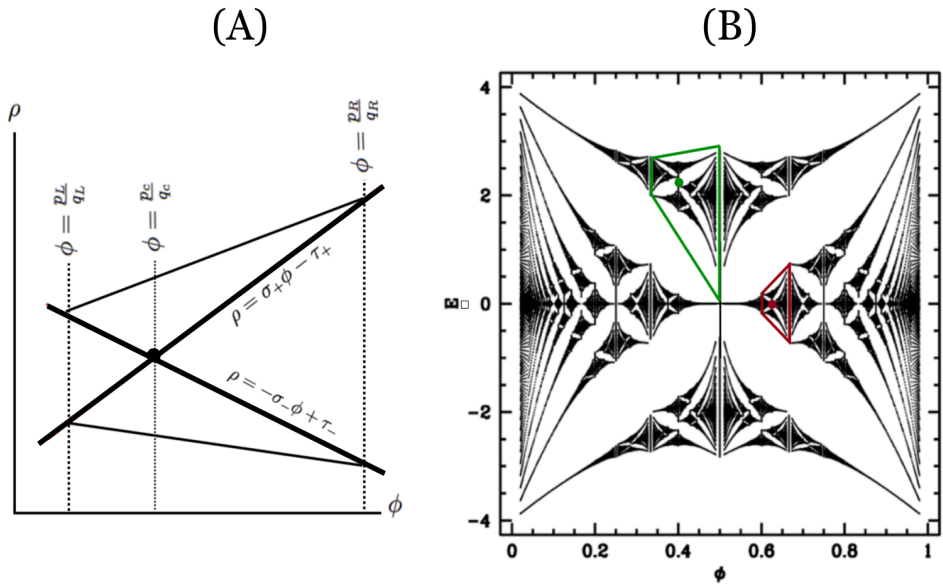}
\leavevmode \caption{  (A) Butterfly Skeleton-  trapelzoidal region,  where the energy gaps of the energy spectrum are simplified down to linear trajectories parametrized as $\rho = \sigma \phi + \tau$.  In the actual butterfly diagram, the linear trajectories become discontinuous. Panel (B) illustrates the relationship between a butterfly and its skeleton with trapezoids enclosing an {\it Edge} butterfly  (green)  and  a {\it Central} butterfly ( red) .  }
\label{sB}
\end{figure}

\section{Butterfly Quadruplets : Labeling a Butterfly}

We now show that  every butterfly in the butterfly graph is uniquely characterized by four integers, christened as the \lq butterfly quadruplets' which are denoted as $Q_B$.

To unveil the number-theoretical designation of butterflies, we represent the two major gaps of a butterfly by two intersecting lines ( See Fig. (\ref{CEnew}) ):
$\rho= \sigma_+ \phi - \tau_+$ and  $\rho= -\sigma_- \phi + \tau_-$ in $(\rho-\phi)$  plane. These linear equations are examples of the Diophantine equation (\ref{DE}) where $\rho= \frac{r}{q}$ is the electron density. 

 The first important point to note is that the $\phi$-coordinate of the  point of intersection of these lines  is $\frac{\tau_+ + \tau_-}{\sigma_+ + \sigma_-}$  which is equal to $\frac{p_c}{q_c}$\cite{book}.
 It is the  flux value at the center of the butterfly. Secondly, once we know the $\phi$ value at the center of the butterfly, Farey relation (\ref{FR}) determines the flux values at the left and the right boundaries as they are the Farey neighbors of $\frac{p_c}{q_c}$.  Two intersecting lines $\rho= \sigma_+ \phi - \tau_+$ and  $\rho= -\sigma_- \phi + \tau_-$  and the two parallel lines $ \phi = \frac{p_L}{q_L}$ and $ \phi = \frac{p_R}{q_R}$ in $(\rho-\phi)$ plane, 
 determine a trapezoidal region.  This  trapezoid is   referred to as the {\it butterfly skeleton}. It is shown in  Fig. (\ref{sB}),
where the four integers $( \sigma_+, \sigma_-, \tau_+, \tau_-)$ uniquely specify a butterfly. The two-dimension at  $(\rho-\phi)$ plane, where every butterfly is represented by the butterfly skeleton,
is the well known Wannier diagram\cite{claro}.

 We now define the butterfly quadruplet as:

\begin{equation}
Q_B=\{\sigma_+, \sigma_-, \tau_++1, \tau_--1 \}
\label{Qb}
\end{equation}
 
We emphasize that in general,  a butterfly triplet $[\frac{p_L}{q_L}, \frac{p_c}{q_c} , \frac{p_R}{q_R}]$ does not label
 a butterfly uniquely as the butterfly graph consists of numerous examples where multiple butterflies share the same flux interval but differ in $(\sigma, \tau)$ quantum numbers. This is shown in
 Fig. (\ref{CEsib}), to be discussed later in section XI.
 
  In view of the butterfly identities listed in Appendix I, for the {\it Central} and the {\it Edge} butterflies, 
 just the pair of integers $( q_c, p_c)$  denoted respectively as $\hat{C}_B$ and $\hat{E}_A$ label the butterflies uniquely. 
 
 For {\it Central}-butterflies:  \\
 $\sigma_+ = \sigma_- =\frac{q_c}{2}, \tau_+ = \frac{p_c-1}{2}, \tau_- = \frac{p_c+1}{2}$,
\begin{equation}
Q_B  =    \{\frac{q_c}{2} ,  \frac{q_c}{2} , \frac{p_c+1}{2} , \frac {p_c -1}{2} \},\,\ \hat{C}_B  =  ( q_c, p_c ); 
\label{Qbc}
\end{equation}

and for {\it Edge}-butterflies:\\
$\sigma_+ =  q_R, \sigma_- = q_L, \tau_+ = p_R-1, \tau_- = p_L+1$
\begin{equation}
Q_B  =   \{q_R,\,\  q_L,\,\ p_R,\,\ p_L \}, \, \  \hat{E}_B  =  ( q_c, p_c ).
\label{Qbe}
\end{equation}
     \begin{figure}[htbp] 
\includegraphics[width = .7 \linewidth,height=.9\linewidth]{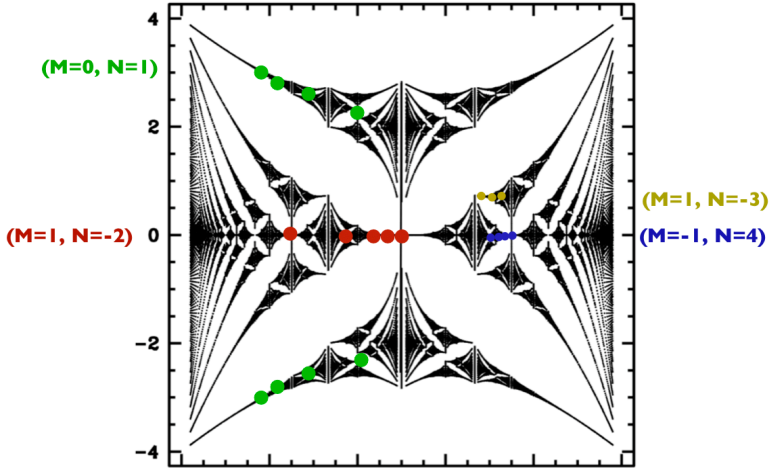}
\leavevmode \caption{ Four  chains of butterflies and the corresponding pair of topological integers $(M,N)$ marked explicitly: $C_{\frac{1}{2}   \rightarrow 0}$ (green),  $C_{0   \rightarrow \frac{1}{2}}$ (red) , 
$C_{\frac{2}{3}   \rightarrow \frac{1}{2}}$ (yellow),  $C_{\frac{2}{3}   \rightarrow \frac{3}{4}}$ (blue ).}
\label{chains}
\end{figure}

 \section{ Chains and Nests of butterflies }
 
\begin{figure}[htbp] 
\includegraphics[width = 1.0\linewidth,height=.8\linewidth]{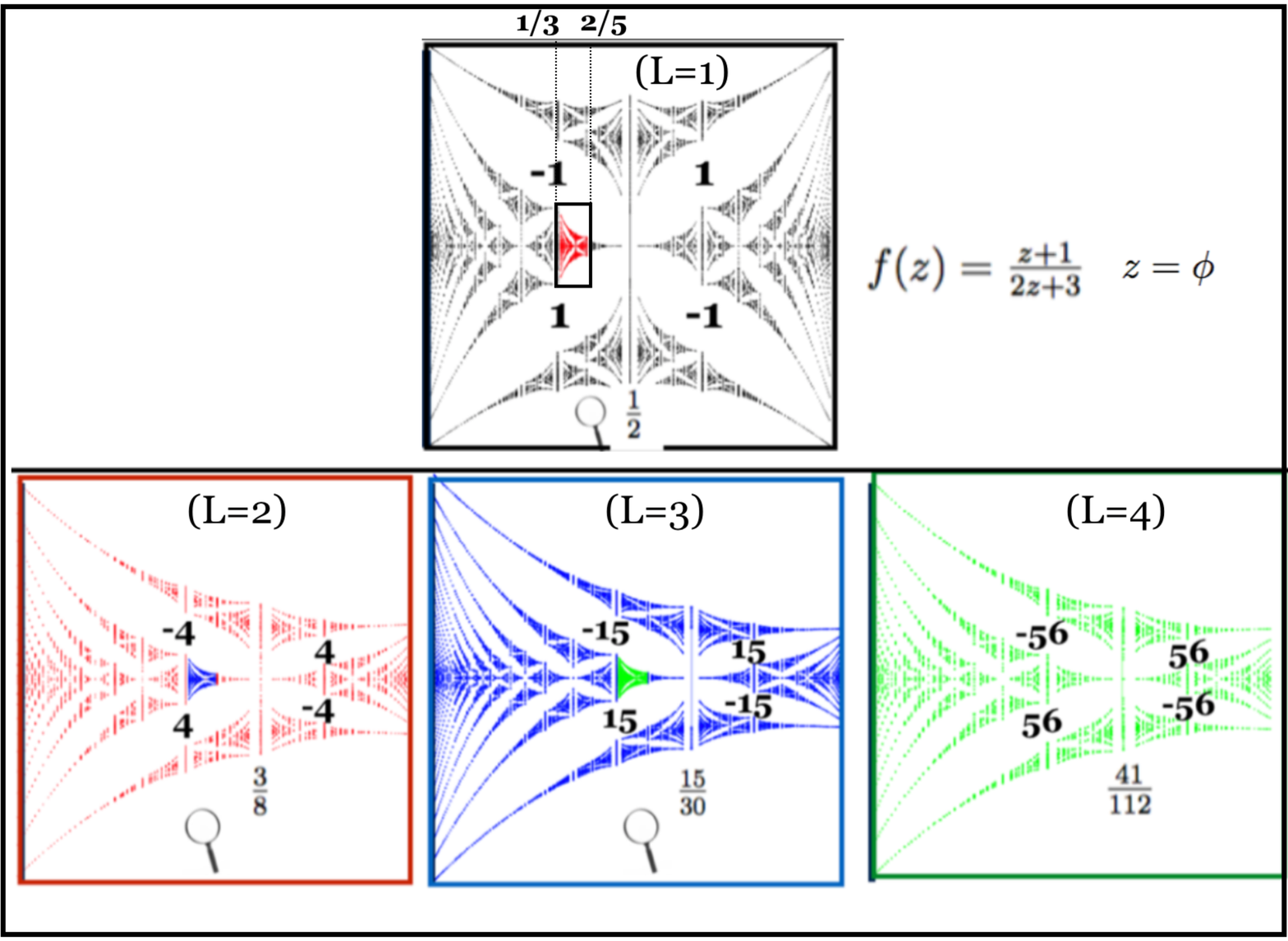}
\leavevmode \caption{ Four levels, $L=1-4$,  of {\it Central}-butterfly  recursions  that results in self-similar images.  Here the root is  the main butterfly:  start with the main 
butterfly and select a sub-butterfly
 (black  box) and zoom into the sub-butterfly maintaining the relation between two successive levels
throughout the nesting.  Major gaps are  labeled with Chern numbers.  The conformal map $f(z)$  that describes the recursive behavior of this hierarchy is also shown. }.
\label{Cs}
\end{figure}

\begin{figure}[htbp] 
\includegraphics[width = 1.0\linewidth,height=.8\linewidth]{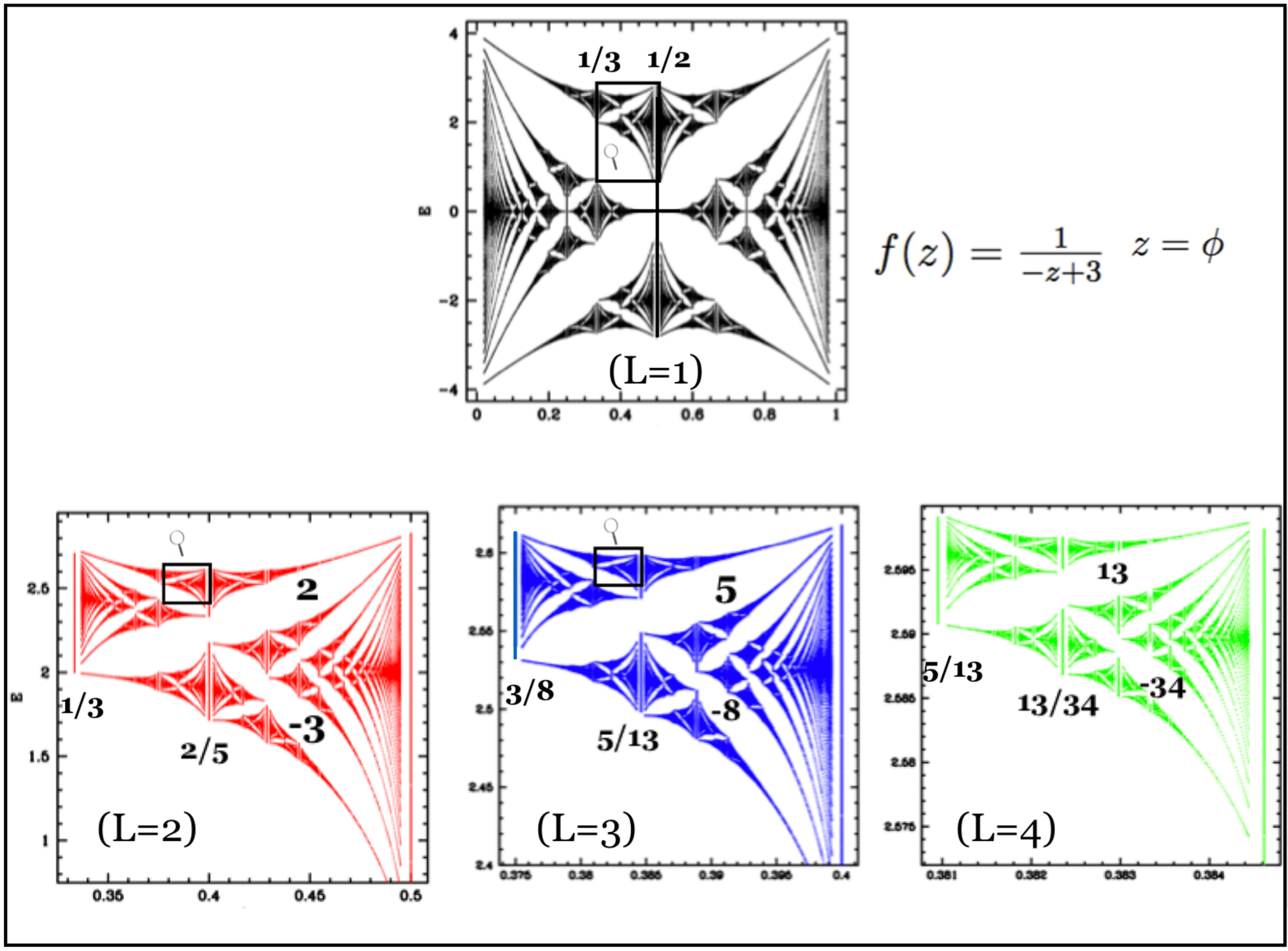}
\leavevmode \caption{ Same as Fig. (\ref{Cs})  for  an {\it Edge}-butterfly. }
\label{Es}
\end{figure}

 The butterfly graph can be viewed as a ``packing" of butterflies arranged in chains, nested  ad infinitum as described in our earlier studies\cite{SW}. A chain of sub-images are given by\cite{SW},

\begin{equation}
\phi_j=\frac{p_0\pm j M_0}{q_0 \mp j N_0},
\end{equation}

where the pair  $(M_0,N_0)$ of quantum numbers  characterize the entire chain.  To construct a chain of  sub-images,
we start with  a band at $\phi_0=p_0/q_0$ and  increase or decrease $\phi$ as shown 
in Fig. (\ref{chains} ) until the spectrum 
of the renormalized Hamiltonian is, once again, a single band. At that  point the second image of the chain  starts and this process continues until the accumulation point of the semi-infinite chain is reached.
Sub-images in the chain are nested and these nesting relationships can be repeated recursively. This is
used to characterize the extent to which the pattern is self-similar as illustrated for the {\it Central} and {\it Edge} butterflies in Figs (\ref{Cs}) and (\ref{Es}) respectively.

The recursive structure of the butterfly fractal, dissected into parts where each sub-image is a microcosm of the entire butterfly graph has been described as  renormalization group trajectories of the
 butterfly integers $(p_L, q_L, p_R, q_R)$,  connecting two successive levels of the hierarchy\cite{SW} . As described in \cite{SW},  we associate a matrix ${\bf B_x(l) }$,  with level $l$ of recursion,
 
 \begin{equation}
  {\bf B}_{\rm x}(l)=
\left(\begin{array}{cc}
q_{\rm x}(l) & p_{\rm x} (l)\cr
-N(l)& M(l) 
\end{array}\right) .
 \end{equation}
 
The recursions can be written as: 
 
   \begin{eqnarray}
 {\bf B}_{\rm x}(l+1)&=&{\bf R}_{\rm x} {\bf B}_{\rm x} (l), 
\label{gen1}
\end{eqnarray}

where the subscript $x = L, R$,  
labels the  transformation matrices $R_L, R_R$  given by, 
\begin{equation}
{\bf R}_{\rm L}=
\left(\begin{array}{cc}
 q^*_{\rm L} &  p^*_{\rm L} \cr
- N^* &  M^*
\end{array}\right),\,\ {\bf R}_{\rm R}=
\left(\begin{array}{cc}
 q^*_{\rm R} &  p^*_{\rm R}- q^*_{\rm R}\cr
- N^* &  M^* + N^*,
\end{array}\right) .
\label{gen2}
\end{equation}
%

\begin{figure}[htbp] 
\includegraphics[width = 1.0\linewidth,height=.8\linewidth]{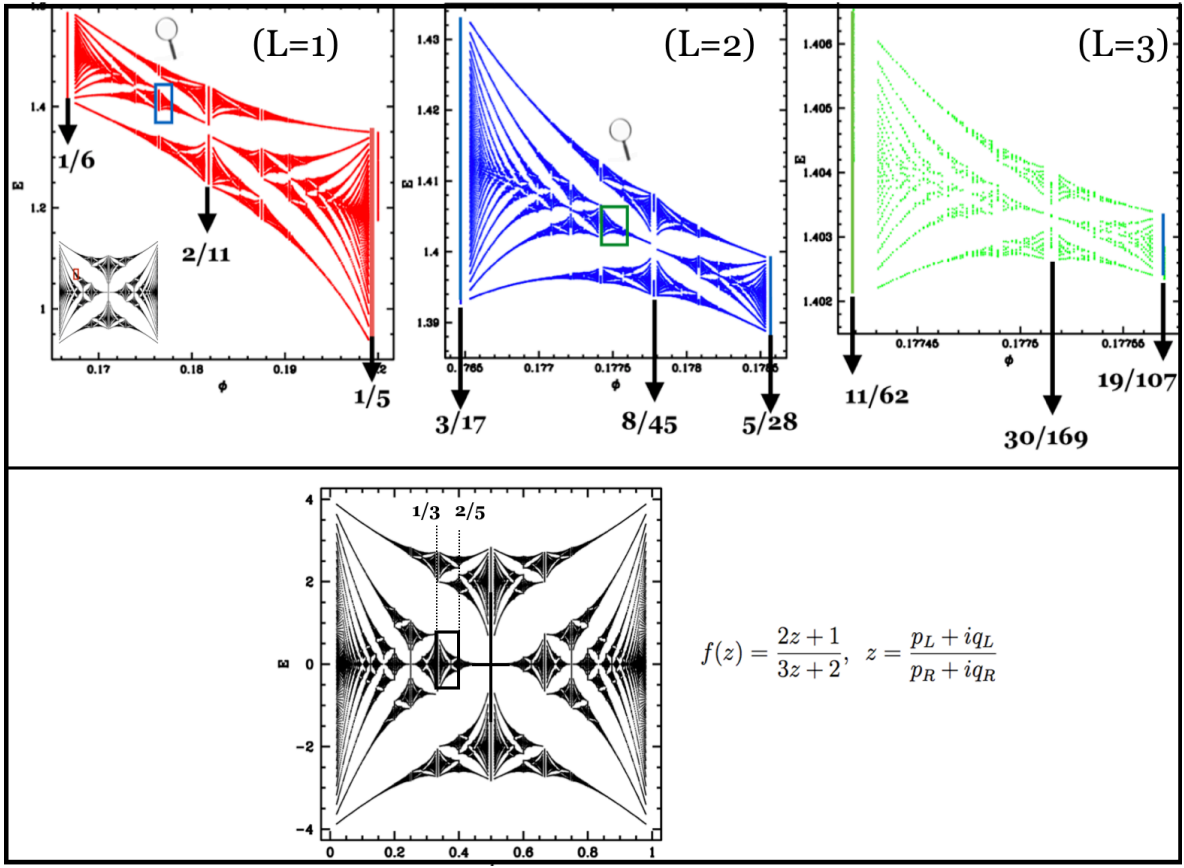}
\leavevmode \caption{ Upper panel shows three level of nesting that results in self-similar images in a C-cell of a  butterfly that lacks any mirror symmetry. This is unlike  Figs.  (\ref{Cs}) and (\ref{Es}) where root is  the main butterfly.  Lower panel
 shows the main butterfly and a sub-image that has same relationship with each other the nested sequences of butterflies in the upper panel. It is the lower panel that determines the constants  of the  transformation.}
\label{Gen}
\end{figure}

The starred integers $( M^*, N^*, p^*_L, q^*_L)$  are fixed by the sub-butterfly
whose nesting structure is being studied as illustrated in the figures (\ref{Cs}), (\ref{Es}) and (\ref{Gen}).  We refer readers to the original paper\cite{SW} for various details of the recursions.

In the special case, where the starting or the root configuration
is the main butterfly, the renormalization equations  can be expressed  in terms of the  renormalization of  a single variable, the magnetic flux $\phi=\frac{p}{q}$, given by

\begin{equation}
\phi(l+1)= \frac{  M^* \phi (l) + p^*_L}{  -N^* \phi (l) + q^*_L}
\label{sim1}
\end{equation}


Figures (\ref{Cs}) and (\ref{Es}) show  butterfly nestings,  described by Eq. (\ref{sim1}), which we will refer as {\it simplified butterfly  recursions}.  An example of  the general case, described by Eq. (\ref{gen1}),
referred as the  {\it general butterfly recursions}, is shown in Fig. (\ref{Gen}).

\section{ Butterfly Recursions as Conformal Transformations}

 We now show that the renormalization trajectories, as described in section $5$  can be expressed as  
conformal transformations\cite{conformal}. This aspect of the recursions will be important in establishing the $\mathcal{ABC}$.

An equation of the form  $w=f(z)= \frac{ a z+b}{cz+d}$ represents a conformal mapping $z \rightarrow w$.  Here  the  constants $(a,b,c,d)$  are  can be determined in terms of 
two sets of triplets:  $(z_1,z_2,z_3)$ and  $(w_1, w_2, w_3)$:

\begin{eqnarray}
 a & = & \det { \left( \begin{array}{ccc} z_{1}w_{1} & w_{1} &1\\ z_{2} w_{2} & w_{2} &1\\  z_{3} w_{3} & w_{3} &1  \end{array}\right) } ,\,\
 b  =  \det {  \left( \begin{array}{ccc} z_{1}w_{1}&z_{1}&w_{1}\\z_{2}w_{2}&z_{2}&w_{2}\\z_{3}w_{3}&z_{3}&w_{3}  \end{array}\right) } \nonumber \\
  c & = & \det {  \left( \begin{array}{ccc} z_{1}&w_{1}&1\\z_{2}&w_{2}&1\\z_{3}&w_{3}&1  \end{array}\right)} ,\,\
 d    =   \det {  \left( \begin{array}{ccc} z_{1}w_{1}&z_{1}&1\\z_{2}w_{2}&z_{2}&1\\z_{3}w_{3}&z_{3}&1  \end{array}\right)}
 \label{abcd}
 \end{eqnarray}
 
We first observe that the equation (\ref{sim1})  is a conformal map  with $z=\phi(l)$ and $w= \phi_{l+1}$ with
the  constants $(a,b,c,d)=(M^*, P^*_L, -N^*, q^*_L)$.  The general case for  butterfly recursions expressed in terms of  two $ 2 \times 2$ matrices described by Eq. (\ref{gen1}) can also be rewritten as a conformal
transformation. By introducing  complex variables $h_x(l)$  where $x=L$ or $R$ as,

\begin{equation}
 h_x(l )= \frac{p_x(l)+ i q_x(l)}{M_l - iN_l},
\label{uv}
\end{equation}

Using Eq. (\ref{gen1}, \ref{gen2}), it is easy to see that the  $h_L(l)$  recursion is a conformal mapping; namely

\begin{equation}
h_L(l+1) = \frac{  q^*_L h_L(l) +  p^*_L}{ (q^*_R-q^*_R ) h_L(l) +  (p^*_R-p^*_L)} ;
\label{hRR}
 \end{equation}

and $h_R$ and $h_c$ satisfy the following equations,

\begin{equation}
h_R(l) = h_L(l)+1,\,\ h_R(l) + h_L(l)  = h_c(l).
\end{equation}

Alternatively, by substituting $M= p_R-p_L$ and $N=q_L-q_R$ ( see Appendix $1$ )  in  Eq. (\ref{gen1}), we obtain a  different  form of recursions where the integers $p$ and $q$  decouple:

\begin{eqnarray}
\left(\begin{array}{c}
p_L(l+1) \cr
p_R(l+1)
\end{array}\right)  & = &  \left(\begin{array}{cc}
q^*_L-p^*_L \,\ & p^*_L\\  
q^*_R-p^*_R \,\ & p^*_R  
\end{array}\right) 
 \left(\begin{array}{c}
p_L(l) \cr
p_R(l)
\end{array}\right) and , \\
\left(\begin{array}{c} 
q_L(l+1) \cr
q_R(+1)
\end{array}\right)   & =  & \left(\begin{array}{cc}
q^*_L-p^*_L \,\ & p^*_L\\  
q^*_R-p^*_R \,\ & p^*_R  
\end{array}\right) 
 \left(\begin{array}{c}
q_L(l)\cr
q_R(l)
\end{array}\right) .
\end{eqnarray}

These equations can be written as  conformal transformations  in terms of  two new variables $e$ and $f$:

\begin{equation}
e_l = \frac{p_L(l)}{p_R(l)},\,\,\ f_l= \frac{q_L(l)}{q_R(l)},
\end{equation}

where $e_l$ and $f_l$ satisfy the following recursions
\begin{eqnarray}
e_{l+1}  & = &  \frac{ (q^*_L-p^*_L)\,  e_l +  p^*_L }{ (q^*_R-p^*_R)  \, e_l+ p^*_R },\,\
f_{l+1}  =  \frac{ (q^*_L-p^*_L)\,  f_l +  p^*_L }{ (q^*_R-p^*_R)  \, e_l+ p^*_R } 
\end{eqnarray}

Finally, we observe that one can express the above two recursions  more elegantly in terms of a complex variable $u$:

\begin{equation}
 u(l) = \frac{p_L(l)+ i q_L(l)}{p_R(l)+i q_R(l)}, \,\ u_{l+1} =  \frac{ (q^*_L-p^*_L)\,  u_l +  p^*_L }{ (q^*_R-p^*_R)  \, u_l+ p^*_R }
 \label{u}
 \end{equation} 
 
 
  \subsection{Geometric origin of Butterfly Recursion}
  
  Butterfly recursions expressed as conformal maps suggests that these recursions may have purely geometric origin.

We now show that Eq. (\ref{sim1}) can  be derived  by mapping  two  sets of butterfly triplets: namely
 $[\phi_L(l), \phi_c(l), \phi_R(l)]$ and 
$[\phi_L(l+1), \phi_c(l+1), \phi_R(l+1)]$, connecting  two successive levels of butterfly nesting. 
 For self-similar hierarchies that start with the main butterfly, one chooses $(z_1,z_2, z_3)=(0, \frac{1}{2},  1)$  and  writes
 $( w_1,w_2, w_3) = ( \phi_L, \phi_C, \phi_R )$.  Using Eq. (\ref{abcd}) along with the Farey relation, Eq. (\ref{FR}), it is found that the constants $(a,b,c,d)$ as:
 $a=p_R-p_L, \,\ b=p_L, \,\ c = q_R-q_L\, \,\ d = q_L$.  This gives $M^*= p_R-p_L$, $-N^*=q_R-q_L$ and $p_L =p^*_L$ and $Q^*_L$.
 in agreement with the Eq. (\ref{sim1}).
 

It is rather remarkable that   Eq. (\ref{sim1}) derived by applying  semi-classical tools to Harper's equation  has such a simple geometric interpretation. It shows that  number theory is intricately
 embedded in the quantum mechanics of the problem. There is no obvious generalization of this type of derivation for the 
 the recursions described by Eq. (\ref{u}), although the conformal property of the map suggests that it is also rooted in geometry and can be  derived without using quantum mechanics. 
 

\subsection{ Self-Similarity }

Scaling factors characterizing self-similar properties of the butterfly graph, determined by the eigenvalues of the transformation matrices $R_L$ or $R_R$ ( Eq. (\ref{gen2}) ) have been discussed in our earlier studies\cite{SW}. Here we note that they are also the eigenvalues of the $ 2 \times 2$ matrix that can be associated with the conformal map that describes the recursions.

Representing the butterfly recursion as  $f(z)= \frac{ b_1 z + b2 }{b_3 z + b_4}$, we can define a 
 $ 2 \times 2$ matrix which we denote as $T_B$

\begin{equation}
 \hat{T}_B= \left(\begin{array}{cc}
b_1 \,\ & b_1\\  
b_3 \,\ & b_4  
\end{array}\right), \, \, \ (b_1 b_4 - b_3 b_2)=1
\label{tb}
\end{equation}

 The eigenvalues of $T_B$, denoted as ( $\zeta, \zeta^{-1}$ ) describe the  asymptotic scaling of integers  $(p_x ,q_x , M, N)$ and the butterfly flux interval $\Delta \phi$:
 \begin{equation}
\lim_{ l \to \infty}  \frac{J(l+1)}{J(l)} = \zeta,\,\,\,\  \lim_{l \to \infty}\frac{\Delta \phi_{l+1}}{\Delta \phi_l}= \frac{1}{\zeta^2}
\ .
\end{equation}

  Here $J$ represents the integers  $(p_x ,q_x , M, N)$.  The scaling factor $\zeta$ are special class of quadratic irrationals\cite{SW} as,

\begin{equation}
 \zeta =\frac{( q^*_L+ M^*)}{2}\pm\sqrt{\left(\frac{q^*_ L+ M^*}{2}\right)^2-1} .
 \label{zeta1}
 \end{equation}
 
Expressed as a  continued fraction expansion, these quadratic irrationals  are given by,
 \begin{equation}
 \zeta= [n^*+1; \overline{ 1, n^*}], \,\ n^* = q^*_L +M^*-2
 \end{equation}

Therefore,  every  self-similar butterfly hierarchy in the butterfly graph
is characterized by a single integer $n^*$.  It is rather remarkable that nature {\it chooses} this special class of quadratic irrationals  to describe self-similar characteristics of the butterfly graph.
For example quadratic irrationals such as silver mean where all entries in the continued fraction  are $2$ are excluded.

We turns out that the C-cell butterflies are characterized by even $n^*$ while for the non-central hierarchies,  $n^*$ can be even or odd. 
\section{ Integral Apollonian Gaskets ( $\mathcal{IAG}$) }


The story of  the $\cal{IAG}$, named after $200$ BC old studies by Apollonian of Perga, begins with four mutually tangent circles such as the ones in Fig. (\ref{pack}). The four curvatures  of these circles
$(\kappa_1, \kappa_2, \kappa_3, \kappa_4)$
 satisfy the following equation, known as the Descartes theorem, found by René Descartes  in 1643\cite{JK2, RF},
\begin{equation}
 2( \kappa_1^2+\kappa_2^2+\kappa_3^2+\kappa_4^2) =  ( \kappa_1+\kappa_2+\kappa_3+\kappa_4)^2 .
 \label{Q4}
 \end{equation}
 The quadratic relation relating four curvatures implies that 
given any three mutually tangent circles of curvatures $(\kappa_1, \kappa_2, \kappa_3)$,  there are exactly two possible circles that are tangent to these  three circles:
\begin{equation}
\kappa_{\pm}=(\kappa_1+\kappa_2+\kappa_3) \pm \sqrt{ \kappa_1 \kappa_2+\kappa_2 \kappa_3+ \kappa_3 \kappa_2}.
\end{equation}
\par\noindent
  Denoting these two solutions as   $\kappa_4, \kappa'_4$, we have
  \begin{equation}
 \kappa_4 + \kappa'_4 = 2 ( \kappa_1+\kappa_2+\kappa_3) .
 \label{l4}
 \end{equation}
Therefore, starting with three mutually tangent circles, we can  construct two distinct quadruplets $(\kappa_1, \kappa_2,\kappa_3, \kappa_4) $ and $(\kappa_1, \kappa_2,\kappa_3, \kappa^{\prime}_4)$. 
As described below, this forms the basis of Apollonian packing described in terms of the Apollonian group.

A remarkable feature of the linear equation  (\ref{l4}) is that  if the original four circles have integer curvature, all of the circles in the packing will have integer curvatures.  

\subsection{ Apollonian Quadruplets}

 In Apollonian packing,  It is natural to consider the  quadruplets  $(\kappa_1, \kappa_2, \kappa_3, \kappa_4)$  rather than the curvatures of individual circles since every circle in the packing is a member of the quadruplet. 
 In other words, the Apollonian packing is packing of Descartes configurations, each characterized by four integers that
satisfy Eq. (\ref{Q4})  and will be referred as the Descartes quadruplets, $Q_D$:

\begin{equation}
Q_D =     \left(\begin{array}{cccc}  \kappa_1,& \kappa_2, & \kappa_3, &\kappa_4 \\ \end{array}\right).
 \label{Qd}
\end{equation}
\par\noindent
 The Descartes quadruplets $Q_D$  are related to another the well known quadruplets, the Lorentz quadruplets $Q_L$:   

\begin{equation}
Q_L =     \left( \begin{array}{ccccc}  N_x , & N_y, & N_z,  & N_t  \\ \end{array}\right),
 \label{Qd}
\end{equation}

where,
\begin{equation}
 N_x^2+N_y^2+N_z^2=N_t^2.
 \end{equation}

\par\noindent
Simple algebraic manipulation  of the Descartes and the Lorentz quadratic forms relates $Q_D$ and $Q_L$. This relationship can be expressed as a matrix equation, 

\begin{eqnarray}
 \left( \begin{array}{c} N_x  \\    N_y  \\  N_z  \\  N_t   \\ \end{array}\right) =  \left( \begin{array}{cccc} 1 & -1 & -1  & -1 \\    0 & 0 & 0 & 2  \\  0 & 1 & -1 & 0\\ 1 & 1 & 2& 1  \\ \end{array}\right) 
  \left( \begin{array}{c} \kappa_1  \\  \kappa_2  \\  \kappa_3 \\  \kappa_4 \\ \end{array}\right) . \end{eqnarray} 
  
The Lorentz quadruplets are generalization of  Pythagorean triplets ( See Appendix $4$ ) and can be constructed from  two Gaussian integers $z_1=m_1+in_1$ and $z_2=m_2+in_2$ which we refer as the
{\it generalized Euclid parameters}.
  \begin{eqnarray}
  N_x  & =  & z_1 z_2^*+ z_1^* z_2 = 2( m_1 m_2 + n_1 n_2),\\
 N_y  & = &  -i(z_1 z_2^*- z_1^* z_2) = 2 ( m_1 n_2 -m_2 n_1),\\
  N_z &  = &  |z_1|^2 - |z_2|^2 = m_1^2+n_1^2-m_2^2- n_2^2,\\
  N_t  & = &  |z_1|^2 + |z_2|^2 = m_1^2+n_1^2+ m_2^2 +  n_2^2.
  \label{EP}
  \end{eqnarray}
  
Therefore,  with every Apollonian we can also associate a quadruplet $Q_A$, christened as the ``Apollonian quadruplet",  
  \begin{equation}
Q_A =     \{ m_1,\,\ m_2,\,\  n_1 ,\,\ n_2  \}
 \label{Qa}
\end{equation}
  
\bigskip\par
  
As described later,  in its simplest form,  Apollonian-Butterfly-connection or  $\mathcal{ABC}$ is essentially a relation between the butterfly quadruplets $Q_B$  and the Apollonian quadruplets $Q_A$ 
 
\section{ Apollonian Packing} 

Starting with a root configuration such as the one shown in  Fig.  (\ref{BP}) or Fig. (\ref{pack}),  the Apollonian packing  is systematically filling in all the empty curvilinear triangular interstices to construct an Apollonian gasket\cite{JK1}. As described below, there are  three equivalent ways to obtain this packing: the  circle inversion, the Apollonian group and  the conformal mapping.
\subsection{ Circle Inversion}

\begin{figure}[htbp] 
\includegraphics[width =  0.8 \linewidth,height=.6 \linewidth]{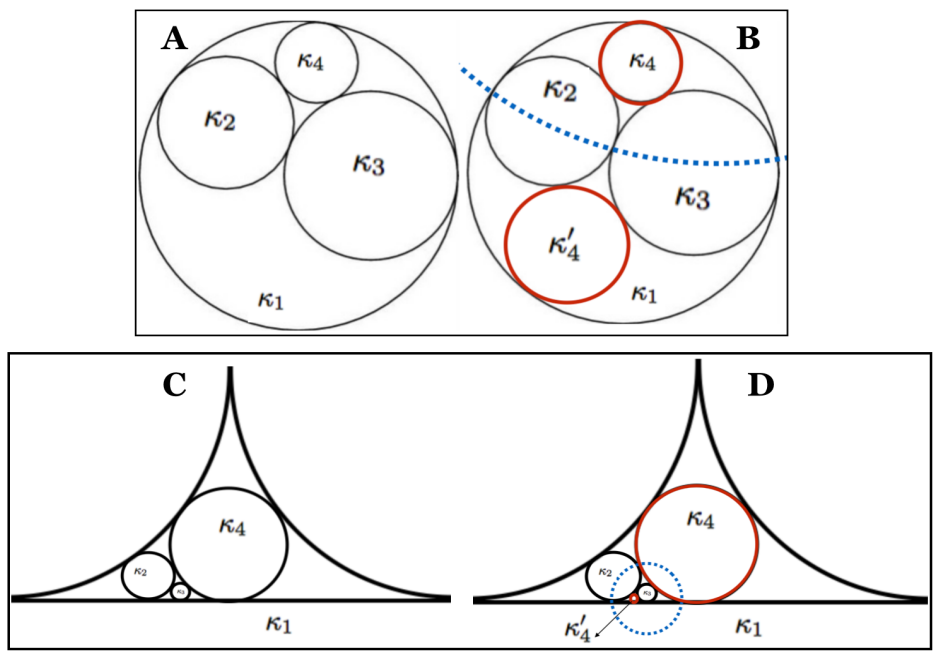}
\leavevmode \caption{  Given a Descartes's configuration of four mutually tangent circles ( A and C), we can form four new configurations by reflecting through a circular mirror ( dotted blue line)  passing through the tangency points of three circles that are shared by the two configurations as shown  on the right ( B and D ).  Here $\kappa_4$ and $\kappa^{\prime}_4$ are mirror images  reflected through the
blue circles.}
\label{pack}
\end{figure}

 Fig.(\ref{pack})   illustrates a geometrical  way to  pack  the Descartes configurations. This simple framework is based on the " circle inversion" that is briefly described in  Appendix II.
Given a Descartes quadruplet $Q_D=(\kappa_1, \kappa_2, \kappa_3, \kappa_4)$, one can obtain another quadruplet $Q^{\prime}_D=(\kappa_1, \kappa_2, \kappa_3, \kappa^{\prime}_4)$ ( See Eq. (\ref{l4}) )
where  $\kappa_4$ and $\kappa^{\prime}_4$ are mirror images about a circular mirror  that passes through the tangency  points of $(\kappa_1, \kappa_2, \kappa_3)$.  Repeating this process where a new Descartes configuration is obtained by circle inversion through any one of the circles 
 provides the simple recipe for Apollonian packing. 
 
 \subsection{The Apollonian Group}

 From the elegant simplicity of plane geometry, we now describe a group theoretical framework to obtain the  Apollonian packing of circles. 
The recursive filling of the space with Descartes configurations can be studied using
Apollonian group. Here  adding additional circles  is accomplished by applying  four matrices $S_i,  ( i = 1-4)$ to a 
a root  configuration of a  Descartes quadruplet. These four matrices are four generators of the group where $S_i^2=1$. This relationship between the circle packing and the Apollonian group follows from Eq. (\ref{l4})   where the generators of the group are,\\
\begin{eqnarray*}
S_1 & =   & \left( \begin{array}{cccc} -1 & 2 & 2  & 2 \\    0 & 1 & 0 & 0  \\  0 & 0 & 1 & 0\\ 0 & 0 & 0& 1  \\ \end{array}\right) ,\,\
S_2= \left( \begin{array}{cccc} -1 & 2 & 2  & 2 \\    0 & 1 & 0 & 0  \\  0 & 0 & 1 & 0\\ 0 & 0 & 0& 1  \\ \end{array}\right), \\
\\
\\
S_3 & =   & \left( \begin{array}{cccc} -1 & 2 & 2  & 2 \\    0 & 1 & 0 & 0  \\  0 & 0 & 1 & 0\\ 0 & 0 & 0& 1  \\ \end{array}\right) ,\,\ S_4=
 \left( \begin{array}{cccc} -1 & 2 & 2  & 2 \\    0 & 1 & 0 & 0  \\  0 & 0 & 1 & 0\\ 0 & 0 & 0& 1  \\ \end{array}\right) .\\
 \label{S4}
\end{eqnarray*}
\
 That is, we view Descartes quadruplets as column vectors $v$ , and the Apollonian group  acts by matrix multiplication, sending $v$  to $Sv$. Note that only one  member of the quadruplet is replaced by this act.  The circle that replaces one of the circles is the mirror image of the replaced circle -- mirror that passes through the tangency points of three circles that are shared by two Descartes configurations. 
 This entire set  of $Q_D$ can be constructed by starting with a root quadruplet and  applying four matrices $S_i$ ad infinitum. 
 
 To have an ordered set of numbers in the quadruplets where the four curvatures appear in monotonically decreasing order, the four matrices $S_i$ are replaced by corresponding matrices which we will represent as $D_i$ so that an ordered ordered set of quadruplets in $v$ transforming to $D v$ produces an ordered set of quadruplets. We represent this relationship between $S_i$ and $D_i$ as:
\begin{equation}
D_i = \hat{O} S_i .
\end{equation}
 
\begin{figure}[htbp] 
\includegraphics[width = 1.0\linewidth,height=.8\linewidth]{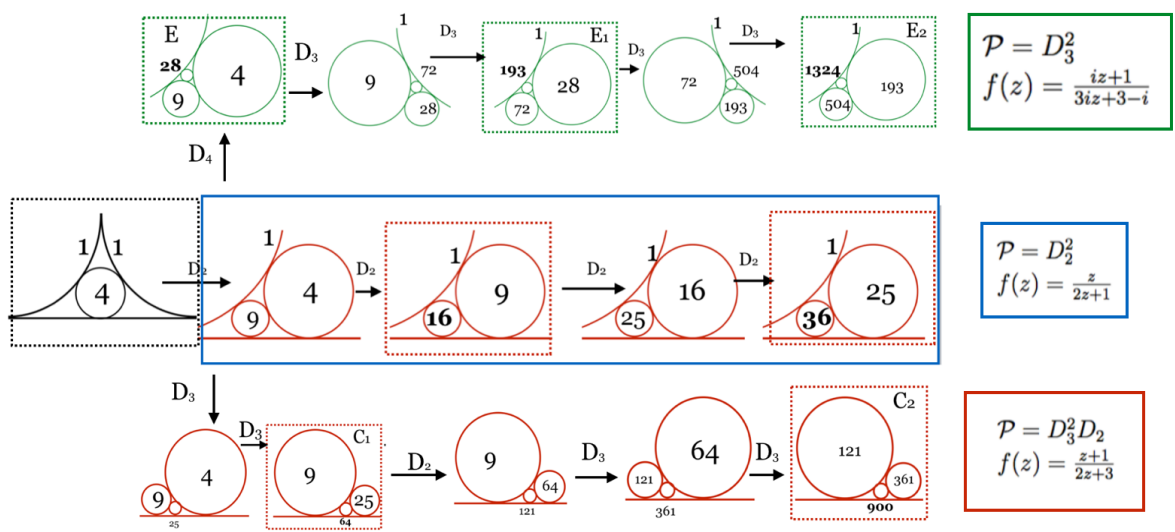}
\leavevmode \caption{ Examples of three chains of packing (circles are not drawn to scale), each characterized by its own string of generators and conformal transformation shown explicitly.
The blue box shows Apollonians  where scaling factor is unity. At the bottom is shown (red circles), a  hierarchy characterized by 
the scaling ratio
$(2+\sqrt{3})^2=[3;\overline{1,2}]$  and at the top (green circles) are nested set of Apollonians with scaling ratio ($ (\frac{3+\sqrt{5}}{2})^2=[2;\overline{1}]$). The dotted boxes correspond to the configurations that represent butterflies as described later. }
\label{AG}
\end{figure}

 Fig. (\ref{AG}) shows three examples of nested sets of Apollonian packings. 
Self-similar configurations are  characterized by  periodic string of the generators of the Apollonian group, denoted as $\mathcal{P}$ in  figure. Eigenvalues of $\mathcal{P}$ determine the scaling of the curvatures.
Figure also shows the corresponding conformal map that is described in the next section below. Note that the eigenvalues of $S_i$ are in general complex. However, the string of operators that describe self-similar configurations have one of the eigenvalues of the form $\zeta^2$ where $\zeta=[ n+1; \overline{1,n} ]$.

 \subsection{ Apollonian Recursions as Conformal Transformations}
 

Any two distinct Descartes configurations can be related  by a conformal map:
\begin{equation}
f(z) = \frac{a_1z+a_2}{a_3 z+a_4}.
\end{equation}
The two sets of triplets  $(z_1, z_2, z_3)$ and $(w_1, w_2, w_3)$  that determine the constants $(a_1, a_2, a_3, a_4)$ are
 three tangency points associated with each Apollonian. 
  For self-similar Apollonian hierarchies, the scaling exponents are the eigenvalues of the $ 2 \times 2$ matrix associated with the map:
 \begin{equation}
\hat{T}_A = \left(\begin{array}{cc}
 a_1 &  a_2 \cr
a_3 &  a_4
\end{array}\right)
\end{equation}

Fig. (\ref{AG}) gives three examples of conformal maps that characterize self-similar hierarchies 
in the Apollonian gasket. As illustrated in  figure,  describing individual hierarchical patterns in terms of $D_i$ or the corresponding conformal transformation is a well defined task.
 However, unlike the packing of the butterflies described by a recursion relation  such as Eq. (\ref{uv}),  there is no single formula that captures
 the nesting of Descartes configurations in the entire packing.

      \begin{figure}[htbp] 
\includegraphics[width = .7 \linewidth,height=.5\linewidth]{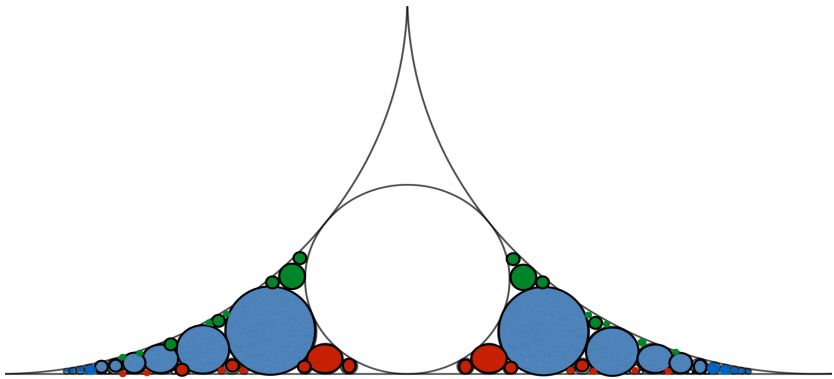}
\leavevmode \caption{Figure shows 
the Papua's chain ( the blue circles)  that  divides the  Apollonian packing into the  {\it Central} (red circles) and the {\it Edge} packing (green circles). }
\label{EC}
\end{figure}
 \section{  Pappu's Chain  and {\it Central} and {\it Edge} division of the Apollonian packing}

 Figure (\ref{EC}) shows the central and the edge packing of circles, (shown with red and green circles respectively) in close analogy with the {\it Central} and the {\it Edge} packing of the butterflies discussed earlier.
 In addition, there are circles that are tangent to both the $x$-axis  and the upper boundary circles, shown in blue in  figure. This "hybrid"  chain of circles, sandwiched between the the central and the edge packing  is an example of  ``Pappu's chain"  investigated by Pappus of Alexandria in the 3rd century AD\cite{Pchain}. 
 
\begin{figure}[htbp] 
\includegraphics[width = .9 \linewidth,height=0.7 \linewidth]{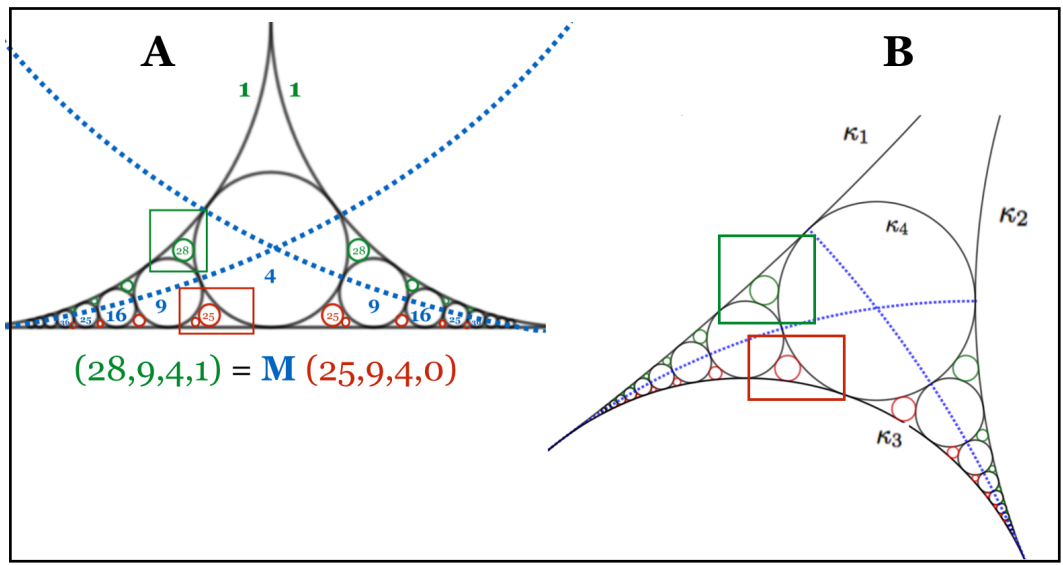}
\leavevmode \caption{ Illustrating the relationship between the {\it Central} and the {\it Edge} packing as the mirror images of each other related by inversion through the circle ( dotted blue shows only section of this circle) that passes through the tangency points of Pappu's chain of circles. The green and the  red boxes show two Descartes configuration that are mirror images, denoted as
 $( 25,4,9,0)= M ( 28,9,4,1)$, described by  the transformation (\ref{cf}).  The panel  B shows
shows the general case where any Apollonian can be  divided
by a Pappu's chain into the central and the edge parts. }
\label{PAPP}
\end{figure}

We note that two consecutive members of the chain are tangent to each other and these tangency points lie on a circle as shown Fig. (\ref{PAPP}).  The key feature of Pappu's chain that is relevant for our studies is the fact that this chain creates two disconnected chains of curvilinear triangles: the upper one we refer as the  ``Edge-chain" ( shown in green) and the lower we refer as the ``Central-chain" ( shown in red). 

The {\it Edge} and the {\it Central}  chains are connected by 
 a conformal transformation, 

\begin{equation}
z \rightarrow f(z) = \frac{z}{-i z+1 }
\label{cf}
\end{equation}
 For the magnetic flux values $\frac{p}{q} > \frac{1}{2}$, the corresponding boundary circle is $|z -(1+\frac{i}{4})|=\frac{1}{4}$  and their curvature is obtained from $\bar{\kappa}_{\frac{p}{q}}$ by $p \rightarrow (q-p) $.
Below we restrict to $\phi \le \frac{1}{2}$. For our discussion below, we will always use the flux values $\phi \le \frac{1}{2}$.

It is easy to see that this transformation maps $x$-axis to the circle $|z-\frac{i}{2}|=1$ and therefore also maps the entire chain of  circles that are tangent to $x$-axis to  chain of circles that are tangent to 
the circle $|z-\frac{i}{2}|=1$.  Under this transformation:
 \begin{equation}
 \frac{p}{q} \rightarrow \frac{pq}{p^2+q^2} + i \frac{p^2}{p^2+q^2}
 \label{pq}
 \end{equation}

 Also the circles that are tangent to x-axis
at  $\phi = \frac{p}{q}$ with curvature $\kappa_ {\frac{p}{q}}$  is transformed to another circle that has curvature $ \bar{\kappa}_{\frac{p}{q}}$:

\begin{equation}
\kappa_ {\frac{p}{q}} =  q^2 \rightarrow \bar{\kappa}_{\frac{p}{q}} = (q^2+p^2-1),
  \end{equation}
  
  Pappu's chain remains invariant under the transformation  (\ref{cf}) as $p=1$ gives $\kappa_{1/q} = \bar{\kappa}_{1/q}$.
  Furthermore, The {\it Central} and the {\it Edge}-packing are mirror images of each other about a circular mirror that passes through mutually  tangency points of Pappu's chain as shown in Fig. (\ref{PAPP}).

     \begin{figure}[htbp] 
\includegraphics[width = .75 \linewidth,height=1.0\linewidth]{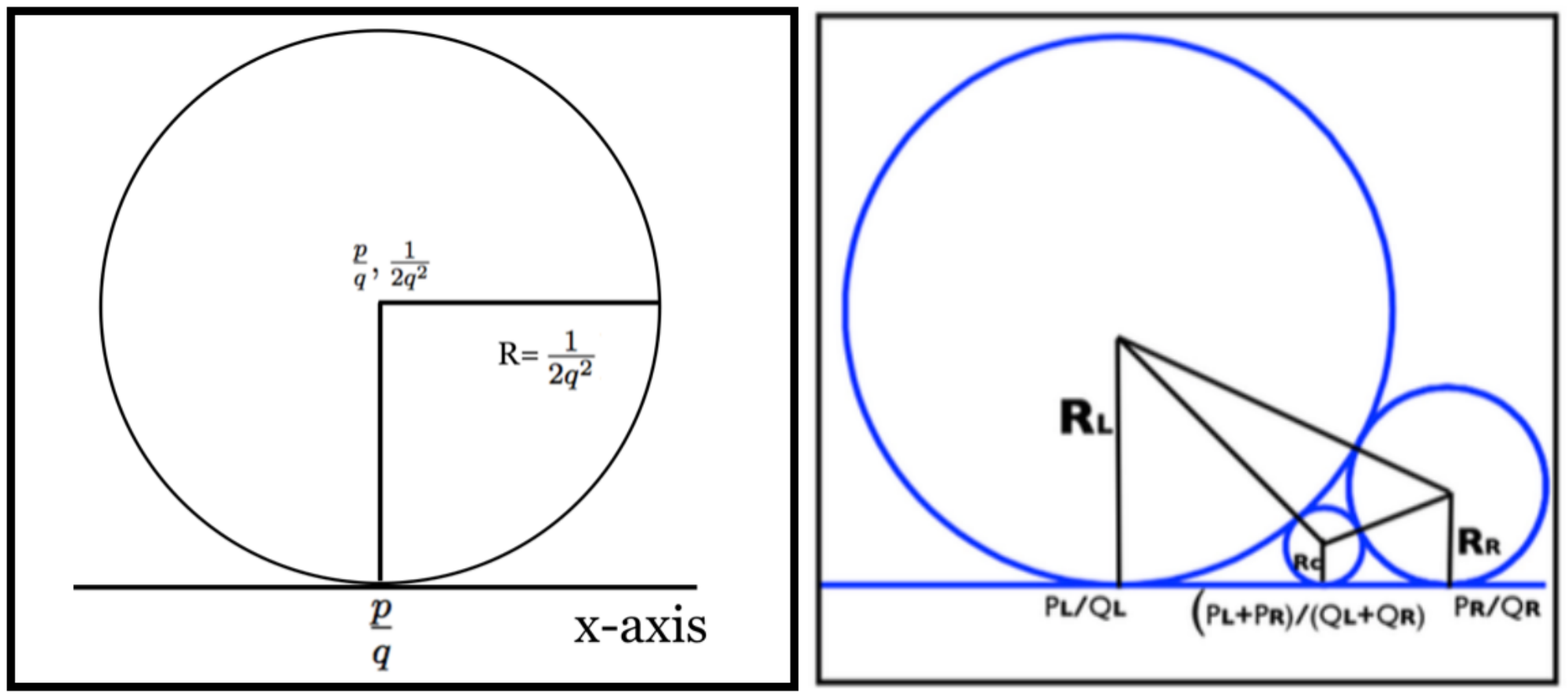}
\leavevmode \caption{ Left panel shows a pictorial representation of a fraction $\frac{p}{q}$ as a circle tangent to x-axis with center at $(\frac{p}{q}, \frac{1}{2q^2})$.  Right panel shows that three Ford circles corresponding
to  three fractions that are Farey neighbors are mutually tangent .}  
\label{F}
\end{figure}
 
\section{Apollonian-Butterfly-Connection ($\cal{ABC}$)}

Butterfly fractal is a {\it packing of  butterflies} where each sub-image is a microcosm of the entire butterfly graph. In this two-dimensional  $(E-\phi)$ landscape,
each sub-butterfly is uniquely labeled by a quadruplet $Q_B$, given by  equation (\ref{Qb}). The $\mathcal{IAG}$ is a packing of four mutually tangent circles  -- the  Descartes configuration. Each configuration
 is uniquely characterized by a set of four integers $Q_A$ ( Eq. (\ref{Qa}) ). We now outline a general framework to show that the butterfly fractal is intimately related to the geometrical fractal $\mathcal{IAG}$.
 This involves  pairing every butterfly  to an  Apollonian and this mapping is one to one. 
  The quantum aspect of the butterfly graph  is reflected in the fact that not all Descartes configurations describe a sub-butterfly in the energy spectrum. However, given a sub-butterfly, one can always find the corresponding Descartes configuration.  The two sets of conformal maps that describe these two fractals are found to be related and the nesting properties of  both fractals are described by the same scaling exponents. 

\subsection{ Ford Circles - Ford Apollonian and {\it Central} Butterflies }

The story of the $\cal{ABC}$  begins with  geometrical representation of fractions in terms of circles, known as  the {\it Ford Circles}.
 Discovered by Lester Ford in $1938$\cite{Ford},  Ford circles provides pictorial representation of fractions. He showed that  every  primitive fraction 
$\frac{p}{q}$ (where $p$ and $q$ are relatively prime)
  can be represented by a circle  in the $x-y$ plan with center at $(\frac{p}{q}, \frac{1}{2q^2})$.  The key characteristic of the Ford circles is the fact that two Ford circles representing two distinct fractions {\it never} intersect . The closest they can come is being tangent to each other and this happens when the two fractions are Farey neighbors. 
  Consequently, with every friendly triplet  $[\frac{p_L}{q_L}, \frac{p_c}{q_c}, \frac{p_R}{q_R}]$,
 one can associate three mutually tangent Ford circles,  that will be nicknamed as the {\it Ford Apollonian}.  
  In this paper, we will scale all curvatures by a factor of two and hence the Ford circles representing a fraction $\frac{p}{q}$ will be represented by a circle of curvature $q^2$. Therefore, Ford Apollonian is characterized by Descartes quadruplet $Q_D=(q_c^2, q_R^2, q_L^2,0)$ and $Q_A=(q_R, q_L, 0, 0)$.

A friendly triplet $[\frac{p_L}{q_L}, \frac{p_c}{q_c}, \frac{p_R}{q_R}]$ that describes a butterfly also describes a Ford Apollonian as the number theory underlying the nesting of butterflies  is related to the packing and nesting of Ford Apollonians. The recursive behavior Ford Apollonians is a mapping of the two  friendly set of triplets $[\frac{p_L}{q_L}(l), \frac{p_c}{q_c}(l), \frac{p_R}{q_R}(l)]$
and $[\frac{p_L}{q_L}(l+1), \frac{p_c}{q_c}(l+1), \frac{p_R}{q_R}(l+1)]$. For self-similar hierarchies such as the one described in Fig. (\ref{ABCc}),
the conformal map to describe the nesting was discussed in section (6.1)  and is given by,

\begin{equation}
\phi_{l+1}= \frac{ a \phi_l+b}{c \phi_l + d} ,\,\ a = (p^*_R-q^*_L), \,\ b = p^*_L, \,\ c = (q^*_R-q^*_l), \,\ d = q^*_L,
\label{fordRR}
\end{equation}

Equation (\ref{sim1}) or equivalently the Eq. (\ref{fordRR})  describes the $\phi$ recursions of all butterflies in the butterfly graph. However, in general, this provides only a partial characterization of butterfly hierarchy.
 As described in section IV,  a friendly triplet $[\frac{p_L}{q_L}, \frac{p_c}{q_c}, \frac{p_R}{q_R}]$
along with additional quantum numbers  such as $(\sigma_+, \tau_+)$
are needed for a unique labeling of a butterfly.  This is because, there are multiple butterflies in the butterfly graph that share same flux interval but differ in $(\sigma, \tau)$ quantum numbers.
 {\it Central} butterflies where $q_c$ is even,
are special class of butterflies that are fully described by  friendly triplets $[\frac{p_L}{q_L}, \frac{p_c}{q_c}, \frac{p_R}{q_R}]$ as the triplets determines the $(\sigma, \tau)$ quantum numbers.

For one to one mapping between the butterflies and the Apollonians, we associate only {\it Central} butterflies with Ford Apollonian. 

As illustrated in Fig. (\ref{ABCc}),
 not every Ford Apollonian represents a butterfly as  butterfly fractal is a quantum fractal that imposes certain symmetry constraints on the butterflies.  For example,
$E \rightarrow -E$ symmetry of the butterfly graph requires $q_c$ to be an even integer for butterflies whose centers lie on the $x$-axis. This is reminiscent of phenomena of missing reflections
of scattering of de Broglie waves in crystalline lattices.

For {\it Central} butterflies and the corresponding Apollonians, we have:

 \begin{figure}[htbp] 
\includegraphics[width = 1.1\linewidth,height=.85\linewidth]{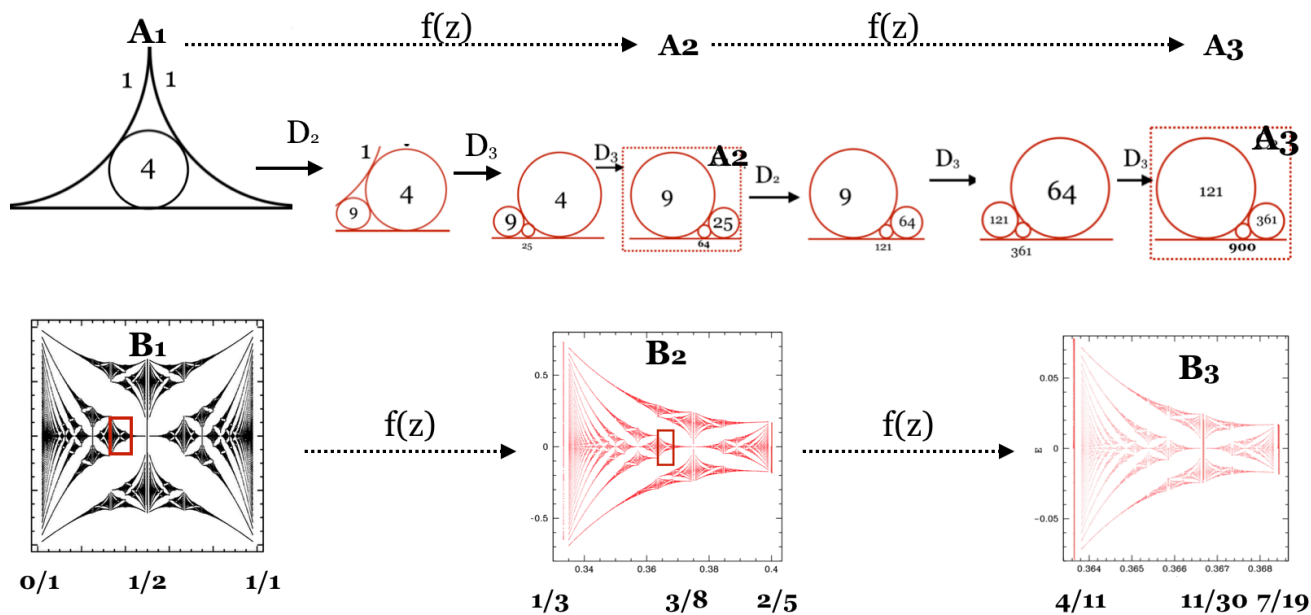}
\leavevmode \caption{ $\mathcal{ABC}$  illustrated. Figure shows nesting for a  {\it Central} Descartes  configurations ( upper panel) where $ (A_1 \rightarrow A_2 \rightarrow A_3) $ 
lead to asymptotically self-similar hierarchy. Lowe panel  shows the  corresponding butterflies:
$ (B_1 \rightarrow B_2 \rightarrow B_3) $. Both are described by the conformal transformation $f(z) = \frac{z+1}{2z+3}$ and scaling  exponent $\zeta= 2+ \sqrt{3}= [3: \overline{1,2}]$. We note that only those Ford Apollonians where the middle circle has even curvature represents a butterfly. }
\label{ABCc}
\end{figure}
\begin{eqnarray}
Q_B & =  & \{ \frac{q_c}{2} ,  \frac{q_c}{2} , \frac{q_L+1}{2}, \frac{q_L-1}{2}  \}, \, \ \hat{C}_B = \{ q_c, q_L \}\\
Q_A & =   & \{ q_R ,\,\ q_L ,\,\ 0 ,\,\  0  \} ,\,\ \hat{C}_A = \{q_R, q_L \}
\label{qbc}
\end{eqnarray} 

In its simplest form, the $\mathcal{ABC}$ for the {\it Central} butterflies is a mapping $\hat{C}_B \rightarrow \hat{C}_A$:

\begin{equation}
\hat{C}_B =   \hat{C}_A \left(\begin{array}{cc}
 1 &  1 \cr
0  &  1
\end{array}\right)
\end{equation}

\underline{Examples}\\

We now   give examples that further illustrate this correspondence. \\

{\bf (1)}  Representing main butterfly with four mutually tangent circles  as shown in Fig. (\ref{ABCc})  is the simplest example of $\mathcal{ABC}$.  Quantitatively, this correspondence
 associates   $Q_B= \{1,1,0,1\}$ to  $Q_A= \{1,1,0,0\}$   as the butterfly is characterized by $\sigma_+=\sigma_-=1$, $\tau_+=0$ and $\tau_-=1$ and the generalized Euclid parameters  for the corresponding Apollonian are $z_1=z_2=1$. Therefore, for the main butterfly,
   \begin{equation}
 \hat{C}_B  =   ( 2,1) , \, \, \
 \hat{C}_A   =   ( 1,1 ).
 \end{equation} 

{\bf (2) } Descartes configurations obtained from circles in the Pappu's chain represent all {\it Central} butterflies with centers at $\frac{p}{q}=\frac{1}{2n}$. These butterflies share the left boundary at $\phi=0$,
and are described by the friendly triplet $ [ \frac{0}{1}, \frac{1}{2n}, \frac{1}{2n-1} ] $. Here $ n=1,2,3....$ represents various butterflies in the chain, with $n=1$ being the main butterfly.

The $\mathcal{ABC}$ for this chain of butterflies can be summed up as,

 \begin{equation}
 \hat{C}_B  =   ( 2n,\,\ 1 ),  \, \, \
 \hat{C}_A   =   ( 2n-1,\,\ 1 )
 \end{equation} 
 
 \begin{equation} 
f(z)= \frac{ z}{ -(2n-2)z + 1} , \, \, \  \hat{T}_B
=  \hat{T}_A =
 \left( \begin{array}{cc}  1\,\ &  0  \\   -(2n-2) \,\  & 1 \end{array}\right) 
 \end{equation}

{\bf (3)} We next consider a chain of {\it Central}-butterflies ( shown in Fig.  (\ref{Cs}) ) : $ [ \frac{n}{2n+1}, \frac{2n+1}{4n+4}, \frac{n+1}{2n+3} ] $\\

 \begin{equation}
 \hat{C}_B  =   ( 4n+4,\,\ 2n+1) , \, \, \
 \hat{C}_A   =  (  2n+3 ,\,\ 2n+1 ), and 
 \end{equation}

 \begin{equation} 
f(z)= \frac{z+n}{2z+2n+1}, \, \  \hat{T}_B
=  \hat{T}_A =
 \left( \begin{array}{cc}  1\,\ &  n  \\   2 \,\  & 2n+1 \end{array}\right) .
 \end{equation}
 
 \bigskip\par
   
 \subsection{ $\mathcal{ABC}$ for  {\it Edge}-butterflies }
 
      \begin{figure}[htbp] 
\includegraphics[width = 0.9 \linewidth,height=0.7\linewidth]{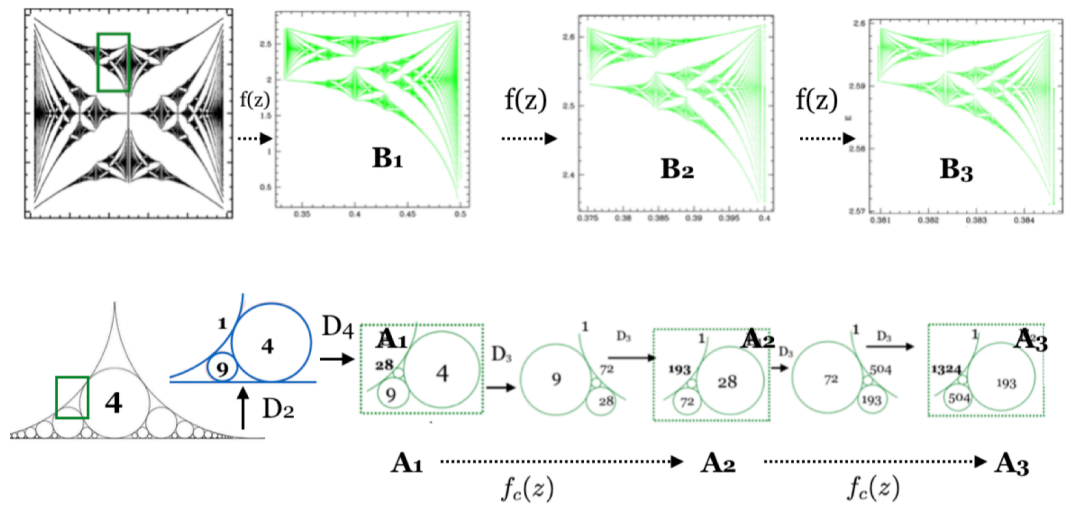}
\leavevmode \caption{  $\mathcal{ABC}$  illustrated for the {\it Edge} butterflies, analog of Fig. (\ref{ABCc}) that describes {\it Central} butterflies. The butterfly and the corresponding Apollonian hierarchies are
characterized by the scaling  exponent $\zeta= \frac{3+ \sqrt{5}}{2}= [2; \overline{1}]$. We note that unlike the self-similar butterfly hierarchy that begins with the main butterfly, the self-similar packing of circles with $\mathcal{P}= D_3^2$ begins with $A_1$.}
\label{ABCe}
\end{figure}

  It is natural to associate the {\it Edge} butterflies with the {\it Edge} packing of Descartes configurations. These configurations are conformal image of {\it Central} packing of circles described by Ford Apollonians.
 
For the {\it Edge} butterflies,  $(\sigma_+, \sigma_-) =  (q_R, q_L)$ and $(\tau_+, \tau_-) = (p_R, p_L)$. And the  generalized Euclid parameters are $z_1=q_R+ i p_R$ and $z_2=q_L + i p_L$. Therefore we have
a special case where the butterfly quadruplets coincide with  the corresponding Apollonian quadruplets:

\begin{equation}
Q_B  = Q_A=\{ q_R ,\,\ q_L ,\,\  p_R ,\,\ p_L  \}
 \label{qbe}
\end{equation}

The nesting characteristics of these {\it Edge}-butterflies are described by the conformal images of the corresponding Ford Apollonians. To illustrate this, we consider the {\it Edge} butterfly,
characterized by the friendly triplet $[\frac{1}{3}, \frac{2}{5}, \frac{1}{2}]$, This is illustrated in   Fig. (\ref{qbe}).
With  three tangency points $( z_1, z_2, z_3)= ( \frac{1}{3}, \frac{2}{5}, \frac{1}{2} )$ of Ford Apollonian and the corresponding $( w_1, w_2, w_3)=( \frac{3+i}{10}, \frac{10+4i}{29}, \frac{2+i}{5})$ 
 obtained from Eq. (\ref{pq}),  Eq. (\ref{abcd}) determines the conformal map $f_c(z)$ as:
 \begin{eqnarray}
 f_c(z) & = &  \frac{iz+1}{3iz+3-i}, \, \ \hat{T}_A =
 \left( \begin{array}{cc}  i \,\ &  1  \\   3 i  \,\  & 3 -i \end{array}\right)
  \end{eqnarray}

The result can be generalized for the entire chain  of {\it Edge} butterflies described by the friendly triplet  $ [ \frac{1}{2n+1}, \frac{2}{2n+3}, \frac{1}{n+1} ] $, $ n=1,2,3...$ The conformal that describes the edge-nesting is given by,

\begin{equation} 
f_c (z) = \frac{ i z + 1} { (n+2)i z + (n+2-i)}, \, \    \hat{T}_A =
 \left( \begin{array}{cc}  i \,\ &  1  \\   (n+2) i  \,\  & n+2 -i \end{array}\right) 
 \end{equation} 
  
  The conformal map that describes the corresponding nesting of the {\it Edge} butterflies butterfly is,
 
  \begin{equation} 
f(z) = \frac{ (n+1)z + 1} { nz + 1}, \, \  \hat{T}_B =
 \left( \begin{array}{cc}  n+1\,\ &  1  \\   n \,\  & 1 \end{array}\right) 
 \end{equation}
 
Although $T_A$ and $T_B$ differ, the two conformal transformations are related as their corresponding matrices have same trace equal to $(n+2)$ and  same the eigenvalues $ E_{\pm} = \frac{n+2  \pm  \sqrt{n^2+4n}}{2}$. 

 The difference in $\hat{T}_A$ and $\hat{T}_B$ for the {\it Edge}-butterflies is due to the fact that renormalization equations describing the butterfly recursions as described in section V
 describe projections of the butterflies on the $\phi$-axis  and do not include the vertical axis of the butterfly.  These transformations involve real variables and it is easy to see that
they describe the nesting of  all Ford-Apollonians , including those that that do not describe {\it Central}-butterflies as they require definite parity for the center  ( even-parity for the denominator q of fraction $\frac{p_c}{q_c}$  and the boundaries ( odd-parity for $q_L$ and $q_R$ ).
These butterfly recursions described earlier  do not distinguish
 two butterflies that share the same flux interval but differ in quantum number $(\sigma, \tau)$. ( See Fig. (\ref{CEsib}), to be discussed later .)  
 
 \subsection{ C-cell and E-cell butterflies without mirror symmetries}
 
\begin{figure}[htbp] 
\includegraphics[width = .8\linewidth,height=.6\linewidth]{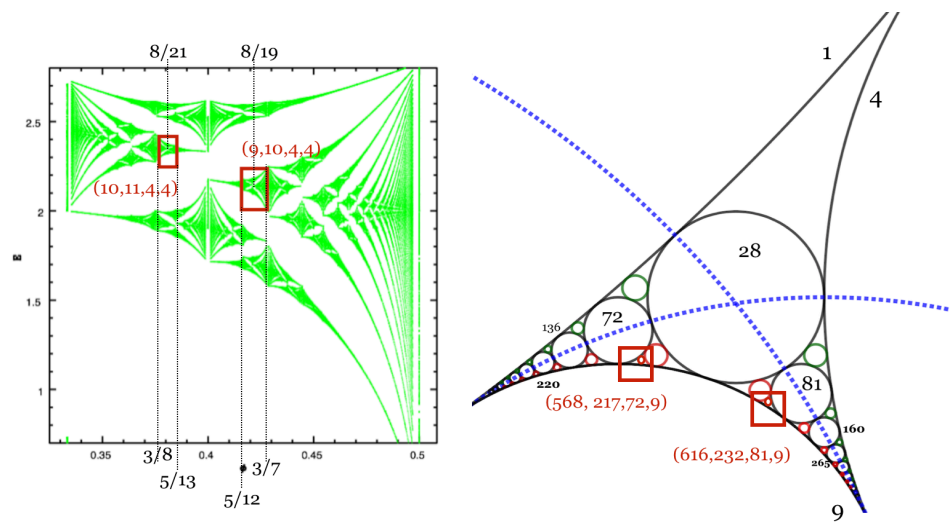}\\
\leavevmode \caption{Left panel shows examples of  butterflies ( red boxes ) that reside in the left and the right C-cell of the  {\it Edge}  butterfly. The flux coordinates of the butterflies are marked. The right panel shows  the corresponding Apollonian.
Here the quadruplets $Q_B$ and $Q_D$ are explicitly marked ( red parenthesis) .} 
\label{C1}
\end{figure}

  Figure (\ref{C1}) gives an example of a  C-cell butterflies and the corresponding Apollonian in the general case where the butterflies do not exhibit mirror symmetry.
  
   Although, explicit formulae to describe $\mathcal{ABC}$ for the general case remains unknown, the task of relating any butterfly to an Apollonian is quite straightforward. 
In other words, given any sub-butterfly, the corresponding Apollonian can be obtained. Below we  give examples that describe this correspondence

The Pappu's chain divides any Apollonian  into C and E cells.  Using elementary but long calculations, we can determine $Q_B$, $Q_D$, $Q_L$ and $Q_A$ for these cases.  For example, for the butterfly in the left C-cell,
$Q_B=(10,11,4,4)$ and the corresponding $Q_A=(15,9,1,0)$, $Q_D=(568,217,72,9)$ and $Q_L=(270,18,145,307)$.

Finally, we note that  the hierarchical nesting properties of C-cell butterflies are encoded in the Pythagorean tree  and has been discussed in earlier studies \cite{Sat18} . This includes both the even parity cases where the butterflies have their center at $E=0$ and the odd parity butterflies whose centers reside of the $E=0$ line.  Appendix IV provides a brief summary of this correspondence.

      \begin{figure}[htbp] 
\includegraphics[width = 0.65\linewidth,height= 0.75\linewidth]{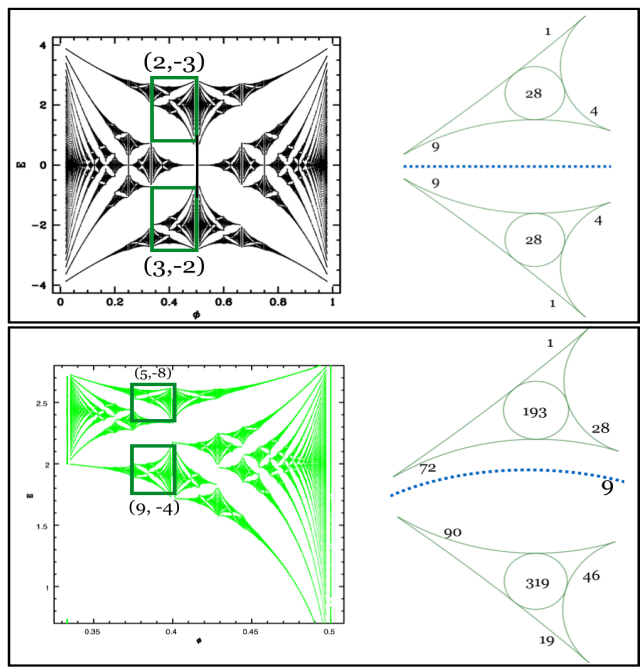} 
\leavevmode \caption{  The left and the right panels respectively show an  upper and the corresponding  lower E-cell butterfly, along with its Chern numbers $(\sigma_+, -\sigma_-)$,
 highlighted with a small box around it. In the top panel,
upper and lower E-cell butterflies are mirror image of each other. 
Bottom  panel shows a generic case without mirror symmetry. 
  In both cases,  the  Apollonian representing the lower edge butterfly is the  the mirror image -- an inversion,  about a circle that is the  base or the lower boundary curve for the C-cell butterflies.}
\label{Eab}
\end{figure}

\subsection{  $\mathcal{ABC}$ and Super-Apollonian Group}

We note that butterfly graph consists of the upper and the lower E-cell butterflies that are in general not related by horizontal mirror symmetry as shown in Fig. (\ref{Eab}).
To obtain  the corresponding Apollonians, we consider 
 we consider an extended root configuration  consisting of a curvilinear triangle and its inversion through the  base or the lower boundary circle representing the C-cell butterflies.
  This process of circle inversion  is mathematically described by an operator $\mathcal{I}=S_1 S_1^{\dagger} S_1$.
We note that it uses $S_i^{\dagger}$ operators which are  generators of the dual Apollonian group as discussed in Appendix. In other words, the entire mathematical framework for
$\mathcal{ABC}$ ``uses"  both the Apollonian and its dual, namely the super-Apollonian group.

   \begin{figure}[htbp] 
\includegraphics[width = .8\linewidth,height=1.125\linewidth]{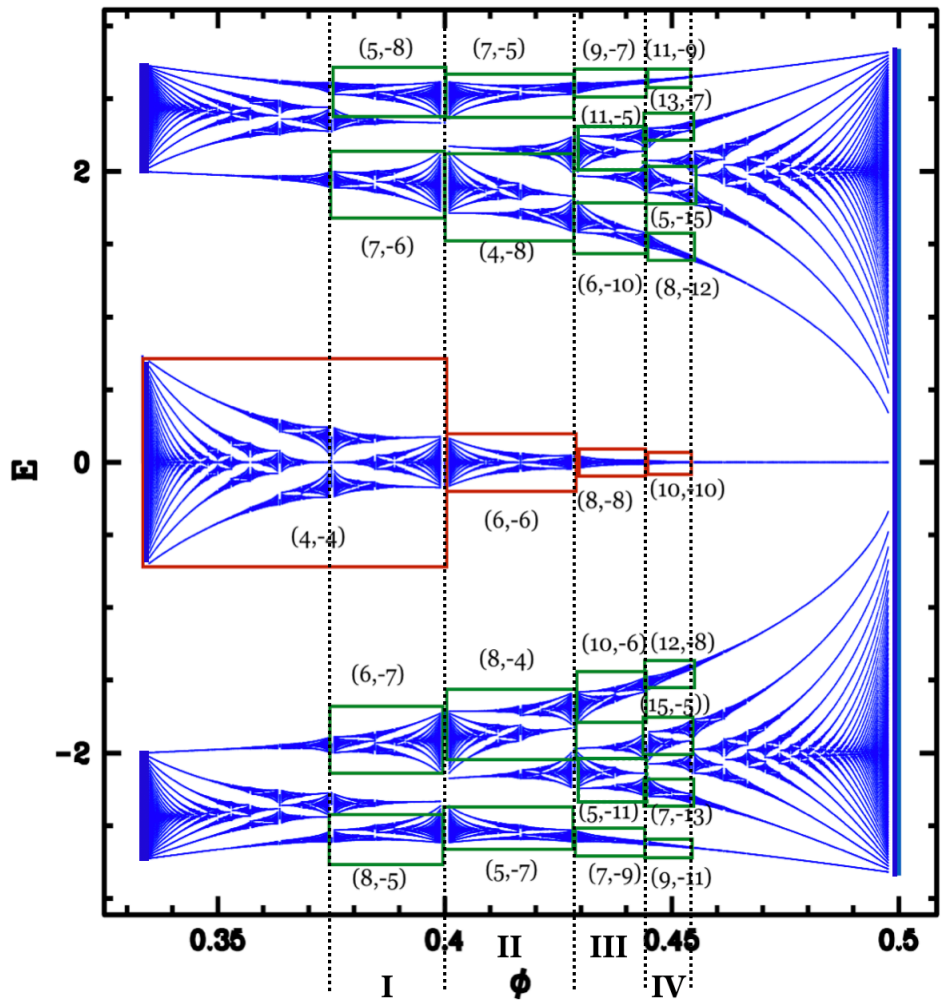}
\leavevmode \caption{  Figure shows four families of {\it sibling}-butterflies (I-IV), stacked vertically (shown within two dotted vertical lines).
 The Central  ( shown in red )  and non-Central ( shown in green)  butterflies in each family share the same flux interval. The two integer numbers near sub-images  show the two Chern numbers $(\sigma_+, -\sigma_-)$ of these butterflies, distinguishing siblings from each other. }
  \label{CEsib}
\end{figure}

   \section{ {\bf Butterfly-Siblings}}

 Figure ~(\ref{CEsib}) shows various examples where a given flux interval hosts multiple butterflies.
Stacked vertically in the butterfly graph, we will refer such a group of butterflies as the {\it butterfly -siblings}. Butterfly siblings differ in $(\sigma, \tau)$ quantum numbers and are characterized by different $Q_B$.
Figure (\ref{CEsib}) shows a chain of {\it Siblings} that share same flux interval while residing both at the center and the edge. The close inspection  shows a very orderly arrangement of the Cherns among the {\it Siblings}.
Mathematical framework underlying $\mathcal{ABC}$ for the {\it Central} and {\it Edge} butterflies assign different $Q_A$ to such siblings.
  
\begin{figure}[htbp] 
\includegraphics[width = .8 \linewidth,height=1.0 \linewidth]{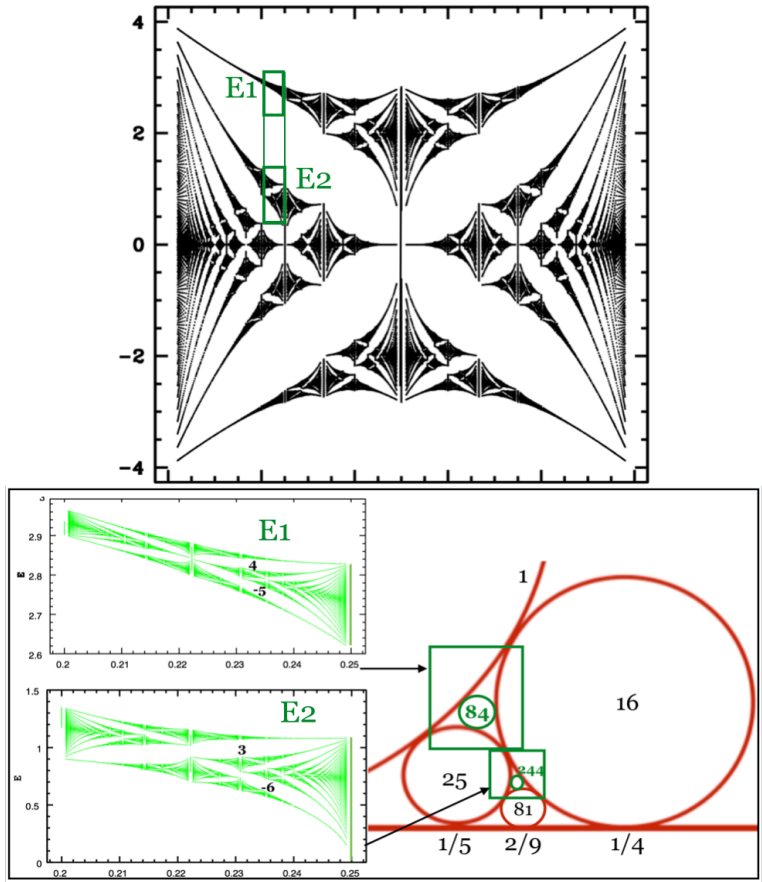}
\leavevmode \caption{ Upper panel marks  two sub-butterflies $E_1$ and $E_2$  that share the same flux interval.  The lower left label shows their blowup along with the Chern numbers and hence are two distinct butterflies represented by two different butterfly quadruplets.  Lower right panel shows the corresponding Descartes configuration.}
\label{EEs}
\end{figure}

One to one mapping between a  butterfly and the corresponding Apollonian  as described above  exhaust all possible scenario  except one involving  butterfly siblings that are both the E-cell butterflies. This case is illustrated in Fig. (\ref{EEs}).  In this case, two distinct E-cell butterflies share the same Pappu's chain and hence the Apollonian mapping of two such butterflies lead to the same Apollonian if we follow the framework described above. 
   To assign two different Descartes configuration to two such siblings, we  use the uppermost triangular space,  which has not been used earlier  by the E or the C-cell butterflies. The process is fully explained and illustrated in the figure (\ref{EEs}).

\section{ Summary and Conclusions} 

Rooted in two competing length scales -- the crystalline lattice and the cyclotron radius, the butterfly spectrum is a marvelous example of a physical incarnation of apparently abstract mathematics. 
A fascinating aspect of the butterfly story comes from recognizing  familiar mathematical  features  lurking the  graph
 such as the Farey relation,  the Pythagorean tree and the Apollonian gasket.  
Relating the spectral landscape to the  Apollonian packing of circles brings out geometrical and number-theoretical
 features pervading the quantum fractal. Fingerprints of quantum effects are reflected in the fact that not all Descartes configurations in the packing
are represented by  butterflies. This is  reminiscent of missing reflections in scattering of  waves from a crystalline solid. 
Number theory dictates the  formation of a self-similar quantum landscape characterized by
quadratic irrationals with  a period-$2$ continued fraction consisting of one and another integer. Emergence of this special class of quadratic irrationals
 is  remarkable. This  further  adds mysticism to this deceptively simple problem of motion of electrons in a crystal  subjected to magnetic field. 
Experimentalists who have seen glimpses of the butterfly in  laboratories\cite{expt} believe that the study of the butterfly offers the possibility of discovering materials with novel exotic properties that are beyond our present imagination. 

The mathematical description of the $\mathcal{ABC}$ as presented here leaves further room for an elegant formulation of the mapping between the two fractals.
Some aspects of this mapping are only empirically understood. The simplicity and richness of the conformal maps such as given by Eq. (\ref{sim1}) and Eq. (\ref{u})  provide solid arguments
for relating  the Central cells of  the butterfly  fractal and the $\mathcal{IAG}$
 is quite appealing. However, they tell only  part of the $\mathcal{ABC}$ story.
Using some empirical results and numerous examples that  exhaust various scenarios from different parts of the spectrum, our discussion makes a convincing case that these two fractals are indeed related. We hope that our studies will stimulate  further investigations
of this problem among mathematicians and physicists.

  
\subsection{ Acknowledgments}. 

It is a great pleasure to thank Richard Friedberg, Jerzy Kocik  and Michael Wilkinson for many stimulating discussions and new insights during the course of this work. Many thanks to Miguel Manacle 
for his help at various stages of investigating $\mathcal{ABC}$. Finally, special thanks to Sam Werner for his comments and proof reading.the manuscript.

\appendix

\subsection{ Butterfly Identities}

Below we list various relationship among the integers $(p_L, q_L, p_R, q_R, p_c, q_c,M, N, \sigma_{\pm}, \tau_{\pm})$ that are   associated with a butterfly.

\begin{itemize}

\item (1)  $N = q_L-q_R$ and $M = p_R-p_L$\cite{SW}.

\item (2)  $\sigma_+ + \sigma_- = q_R+q_L=q_c$ and $\tau_+  +  \tau_- = p_R+p_L=p_c$\cite{book}.

\item  (3) For the {\it Central butterflies}, $\sigma_+ = \sigma_- = \frac{q_c}{2} $, $p_c=q_L$ , $\tau_+ = \frac{p_c-1}{2}$ and $\tau_- = \frac{p_c +1}{2}$\cite{book}.

\item (4) For the upper {\it Edge butterflies},  $\sigma_+ = q_R$ , $\sigma_- = q_L$, $\tau_+ = p_R-1$ and $\tau_-= p_L+1$.
For the lower edge,  $\sigma_+ = q_L$ , $\sigma_- = q_R$, $\tau_+ = p_L+1$ and $\tau_-= p_R-1$

 Butterfly identities $(1-3)$ have been proven earlier\cite{book, SW} . To prove (4) , we note that for the {\it Edge butterflies}, the butterfly center consists of two bands separated by a gap, splitting the quantum number $N$ of the butterfly  into $N_1$ and $N_2$ where $N = N_1+N_2$. Here $N_1$ is the outer band and $N_2$ is the inner band. But $N_1 = q_L-(q_R+q_L)= -q_R$. Now,  $\sigma_+ = 0-N_1= q_R$. Therefore $\sigma_- =  q_L$.

Similarly, it can be shown that $\tau_+ = p_R-1$ and $\tau_- = p_L+1$. 
\end{itemize}
 
  \subsection{ Circle Inversion}
   
 Inversion in a circle is a method to convert geometric figures into other geometric figures. There is a fairly easy mathematical relationship between a figure and its  its inversion.
 The geometric tool of inversion in a circle is often used in mathematics to simplify solving some problems.
These transformations preserve angles and map generalized circles into generalized circles, where a generalized circle means either a circle or a line (loosely speaking, a circle with infinite radius). Many difficult problems in geometry become much more tractable when an inversion is applied. 
 
 To invert a number in arithmetic usually means to ``take its reciprocal". A closely related idea in geometry is that of ``inverting" a point. 
 In the plane, the inverse of a point $P$ with respect to a reference circle C with center O and radius $r$ is a point $P^{\prime}$ , lying on the ray from O through P such that

 \begin{equation}
OP\times OP^{\prime} =r^2.
\end{equation}

If we wish to invert a more complex figure than a single point, we simply invert every point in the figure and the resulting set of points becomes the inverted figure.

A key property of inversion is the following: If every point on a ``circle" is inverted through a circle C , the result will be a ``circle".

One can obtain the curvatures  of the conformal images, using the inversion formula,
\begin{equation}
r^{\prime} = r \frac{R^2}{d^2-r^2},
\end{equation}

where $r^{\prime}$, $r$ and $R$ respectively are the radii of the image, the object and the mirror and $d$ is the distance between the centers of the object ( center at $( \frac{p}{q}, \frac{1}{2q^2})$ )
 and the mirror ( center at $(0,1)$ ). 
This gives $d^2 -r^2 = \frac{p^2+q^2-1}{q^2}$ and the curvature $\kappa^{\prime} = 2(p^2+q^2-1)$.
 
 In case where the object and the
mirror circle touch, $d= r+R$ and we obtain,
\begin{equation}
\kappa^{\prime} = \kappa + \frac{2}{R}
\end{equation}

\begin{figure}[htbp] 
\includegraphics[width = 0.7 \linewidth,height=.55\linewidth]{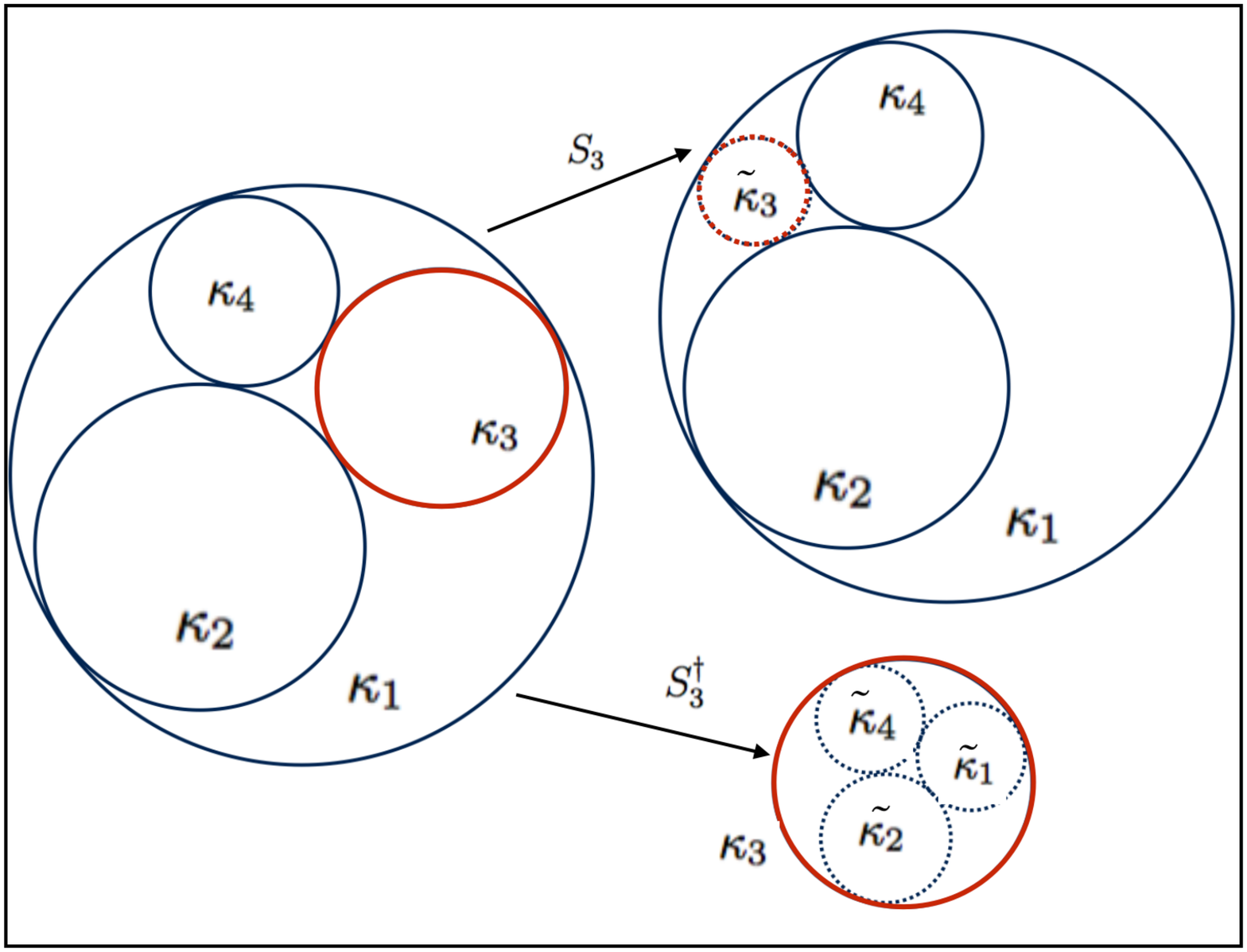}
\leavevmode \caption{  Illustrating the actions of $S_3$ and $S_3^{\dagger}$ on a Descartes configuration}
\label{SSA}
\end{figure}

   \begin{figure}[htbp] 
 \includegraphics[width = 0.9\linewidth,height=.7\linewidth]{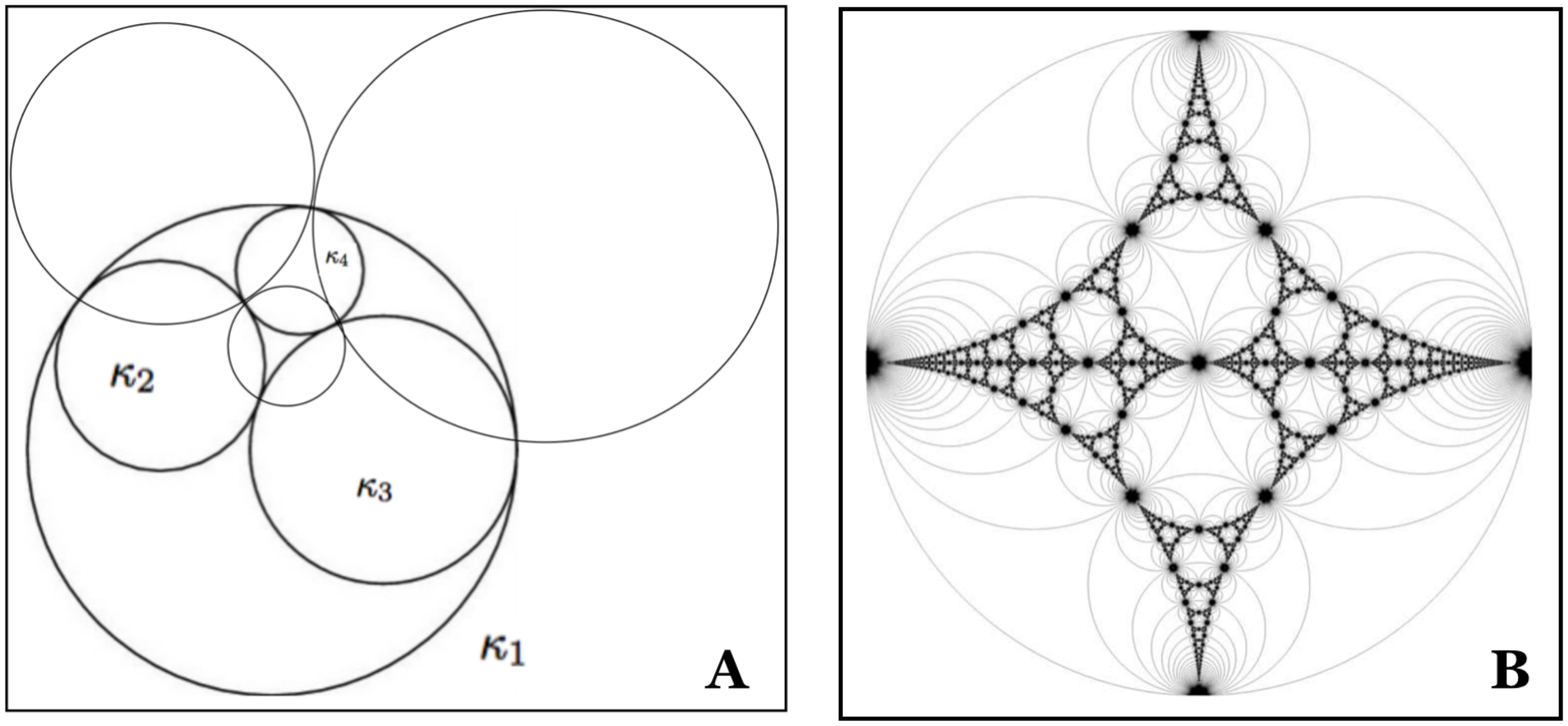}
\leavevmode \caption{ A: Descartes configuration ( dark circles) and its dual ( lighter circles). B: shows  both the Apollonian packing ( dark circles) and its dual ( lighter circles). }
\label{dual1}
\end{figure}

\subsection{Dual and Adjoint -- Super Apollonian Group\cite{Apacking} }
  
  Given a Descartes quadruplets $Q_D=(\kappa_1, \kappa_2,\kappa_3,\kappa_4)$, there exists another Descartes configuration $\bar{Q}_D = (\bar{\kappa}_1,\bar{ \kappa}_2,\bar{\kappa}_3, \bar{\kappa_4})$. Geometrically the $Q_d$ and its dual $\bar{Q}_d$ are related as follows: 
  Given a Descartes configuration, we can obtain a related  {\it dual} configuration by drawing circles through three of the tangency points of the four circles. Since there are four such trios in any quadruple, this construction gives us four new mutually tangent circles and we will denote this dual configuration as $\bar{Q}_d$. In other words $\bar{Q}_d$ is obtained from $Q_d$ consisting of the four circles each of which passes through the three tangency points avoiding one circle.  These two {\it dual} configurations are related by an operator $U$:
  
 \begin{eqnarray}
\left(   \begin{array}{c}  \bar{\kappa}_1 \\  \bar{\kappa}_2 \\  \bar{\kappa}_3 \\  \bar{ \kappa}_4 \end{array}\right)  =
   \frac{1}{2}  \left( \begin{array}{cccc } -1 & 1 & 1 & 1 \\    1 & -1   & 1 & 1  \\  1 & 1   & -1 & 1\\ 1 & 1 & 1 & -1 \\ \end{array}\right)   \left(  \begin{array}{c} \kappa_1 \\  \kappa_2 \\  \kappa_3  \\  \kappa_4   \\ \end{array}\right) \equiv  U \left(  \begin{array}{c} \kappa_1 \\  \kappa_2 \\  \kappa_3  \\  \kappa_4   \\ \end{array}\right) 
 \label{abc1}
\end{eqnarray}

Figure \ref{dual1} shows the relationship between $Q_D$ and its dual $\bar{Q}_D$ along with the Apollonian packing and  the corresponding dual packing.
The Apollonian group  is generated by four inversions w.r.t. those dual circles.

With the four operators $S_i$, we next define a second set of four operations $S_i^{\dagger}$ that  correspond to inversion in one of the four circles in a Descartes configuration.  Fig. (\ref{SSA}) compares and contrasts
$S_i$ and $S^{\dagger}_i$ using $S_3$ and $S_3^{\dagger}$.  $S_3$  changes only $\kappa_3$ while with $S_3^{\dagger}$,  $\kappa_3$ remains fixed while  curvatures of the other three circles change.

Given an Apollonian group $\cal{A}$ with four generators $(S_1, S_2, S_3, S_4)$, we can define a dual Apollonian group $\bar{\cal{A}}$ with generators $(S^{\dagger}_1, S^{\dagger}_2, S^{\dagger}_3, S^{\dagger}_4)$, where
$S^{\dagger}_i$ is the transpose of $S_i$. The two groups $\cal{A}$ and $\bar{\cal{A}}$ are  related as their generators satisfy the following equation. 

\begin{equation}
S^{\dagger}_i = U S_i U
\end{equation}

By combining $\cal{A}$ and $\bar{\cal{A}}$,  one can form a new group, known as the {\it super Apollonian Group} $\cal{A}^S$ with eight generators,
$( S_1, S_2, S_3, S_4, S^{\dagger}_1, S^{\dagger}_2, S^{\dagger}_3, S^{\dagger}_4 )$.
The curvature of the dual circle  is $\bar{\kappa} = \kappa_1 \kappa_2 +  \kappa_2 \kappa_3 +  \kappa_3 \kappa_1$, where $\kappa_i$ ($ i=1,2,3$) are the curvatures of the 3 circles whose tangency points define the dual circle.

To have an ordered set of numbers in the quadruplets where the four curvatures appear in monotonically decreasing order, the four matrices $S_i$ are replaced by corresponding matrices which we will represent as $D_i$ so that an ordered $v$ transforming to $D v$ produces an ordered set of quadruplets: $D_i = \hat{O} S_i$.\\

\subsection{ Pythagorean Tree  }

    \begin{figure}[htbp] 
\includegraphics[width = .7 \linewidth,height=.6\linewidth]{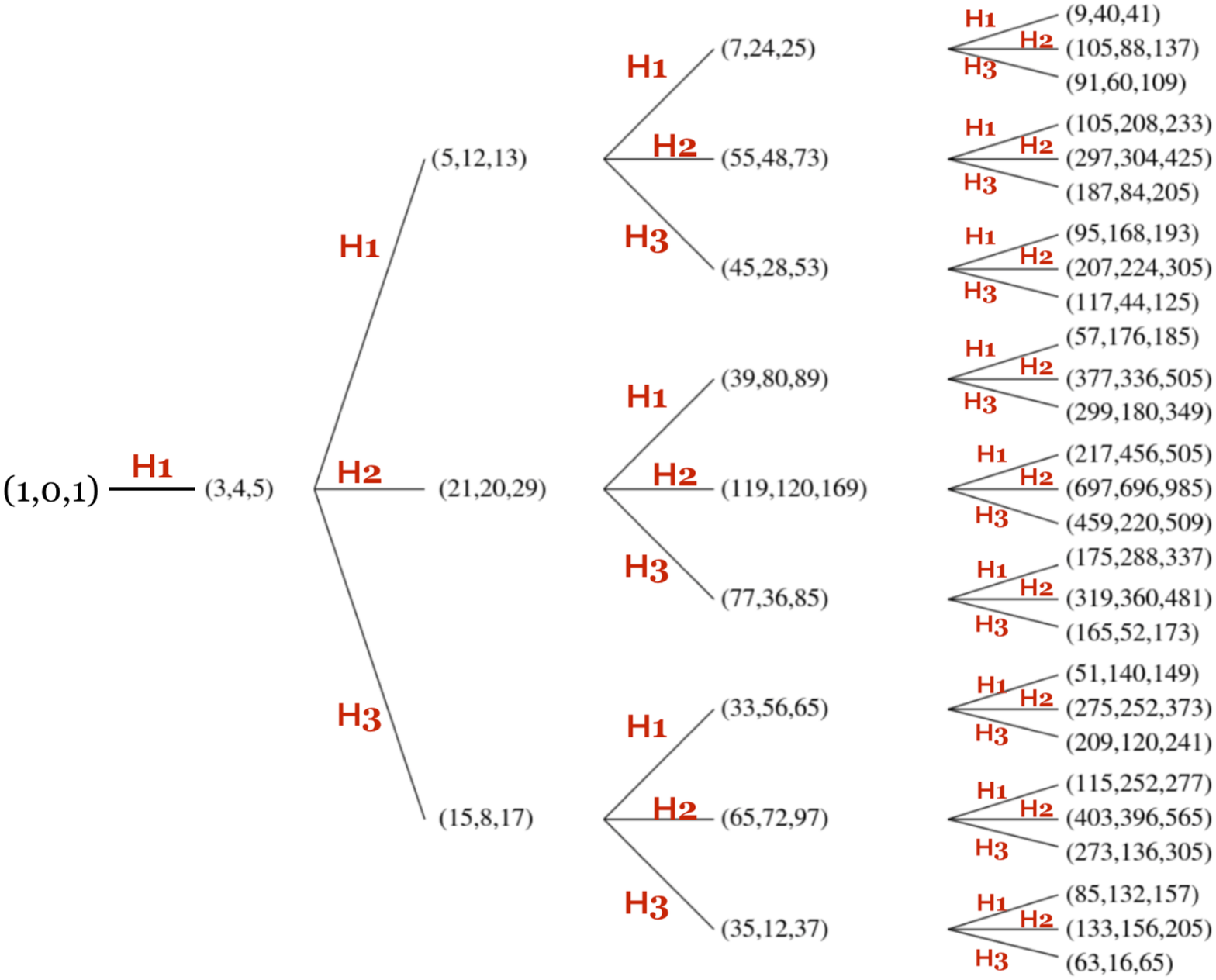}
\leavevmode \caption{ Pythagorean tree encodes nesting for all C-cell butterflies}
\label{PT}
\end{figure}

A Pythagorean triple is a set of three positive integers $(a, b, c)$   having the property that they can be respectively the two legs and the hypotenuse of a right triangle, thus satisfying the equation,

\begin{equation}
 a^2+b^2=c^2.
 \label{P}
 \end{equation}
 The triplet is said to be primitive if and only if a, b, and c share no common divisor.

In $1934$,  B. Berggren  discovered\cite{Sat18}  that the set of all primitive Pythagorean triples has the structure of a rooted tree. In 1963,  this was rediscovered by the  Dutch  mathematician  F.J.M.  Barning   and  seven years later by  A. Hall\cite{Hall}  independently.  Using algebraic means, it was shown  that all primitive Pythagorean triplets can be generated by three matrices  which we label as $H_1$, $H_2$ and $H_3$ as shown in Fig. (\ref{PT}). The three matrices are given by,

\begin{equation}
H_1=\left( \begin{array}{ccc} 1 & -2 & 2  \\    2 & -1 & 2  \\  2 & -2  & 3 \\ \end{array}\right),
H_2=\left( \begin{array}{ccc} 1 & 2 & 2  \\    2 & 1 & 2  \\  2 & 2  & 3 \\ \end{array}\right),
H_3=\left( \begin{array}{ccc} -1 & 2 & 2  \\    -2 & 1 & 2  \\  -2 & 2  & 3 \\ \end{array}\right) 
\label{H3}
\end{equation}

As discussed earlier, the Descartes's quadruplets  $Q_D$ are related to the Lorentz quadruplets $Q_L$.  This provides the direct relation between $H_1, H_2, H-3$ and the generators $D_i$
of the Apollonian group:

For butterflies with their centers at $E=0$ ,  $q_c$ is always even.  For such {\it even- parity} butterflies,  the corresponding Pythagorean triplets  are:
 
  \begin{eqnarray}
  (n_x,n_y,n_t) & = &  ( q_L q_R, \frac{1}{2}(q_R^2-q_L^2),  \frac{1}{2}(q_R^2+q_L^2))
  \label{bspinor}
  \end{eqnarray} 

For  C-cell butterflies whose center do not reside at $E=0$,  $q_c$ is odd. Such  {\it odd-parity} butterflies can be described by a ``dual" Pythagorean tree --
a tree where the legs  of the right triangle ( that is $n_x$ and $n_y$ ) are interchanged. The Pythagorean triplets for these butterflies where $q_R$ and $q_L$ have opposite parity are obtained by multiplying by a factor of $2$ in Eq. (\ref{bspinor}).

 We note that the  integers $n_x$ and $\kappa_0$ determine  the horizontal size $\Delta \phi$ of a butterfly:
$\Delta \phi = |\frac{p_R}{q_R}-\frac{p_L}{q_L}| = \frac{1}{q_L q_R} =   \frac{1}{n_x}$.

In view of the fact that the Pythagorean tree or its dual preserve the parity,  the hierarchical character of the E-cell butterflies cannot be described by the Pythagorean tree. Perhaps a quadruplet tree (a  tree of Lorentz quadruplets ) can be constructed as a generalization of the Pythagorean triplet tree. However, to best of our knowledge, there is no proper generalization of the Pythagorean tree to the quadruplet tree although the topic has been the subject of some discussion in the literature\cite{JK}.  We note that  some of the Lorentz quadruplets that describe the E-cell butterflies are
are non-primitive and therefore the existence of any tree structure that describes the entire butterfly graph seems rather unlikely.


\end{document}